\documentclass[twocolumn,showpacs,aps,pre,floatfix,superscriptaddress]{revtex4-1}
\usepackage{amsmath,amssymb,eucal}
\usepackage{graphicx}
\usepackage{float}
\usepackage{multirow}
\begin{document}
\title{Zero-Temperature Relaxation of Three-Dimensional Ising Ferromagnets}
\author{J. Olejarz, P.~L.~Krapivsky, and S.~Redner}

\affiliation{Center for Polymer Studies,
and Department of Physics, Boston University, Boston, MA~ 02215, USA}

\begin{abstract}
  We investigate the properties of the Ising-Glauber model on a periodic
  cubic lattice of linear dimension $L$ after a quench to zero temperature.
  The resulting evolution is extremely slow, with long periods of wandering
  on constant energy plateaux, punctuated by occasional energy-decreasing
  spin-flip events.  The characteristic time scale $\tau$ for this relaxation
  grows exponentially with the system size; we provide a heuristic and
  numerical evidence that $\tau\sim \exp(L^2)$.  For all but the smallest-size
  systems, the long-time state is almost never static.  Instead the system
  contains a small number of ``blinker'' spins that continue to flip forever
  with no energy cost.  Thus the system wanders {\em ad infinitum} on a
  connected set of equal-energy blinker states.  These states are composed of
  two topologically complex interwoven domains of opposite phases.  The
  average genus $g_L$ of the domains scales as $L^\gamma$, with
  $\gamma\approx 1.7$; thus domains typically have many holes, leading to a
  ``plumber's nightmare'' geometry.

\end{abstract}
\maketitle

\section{Introduction}

We study the evolution of the homogeneous Ising ferromagnet on a periodic
cubic lattice in which the spins are endowed with zero-temperature Glauber
dynamics.  Starting from an initial state of zero magnetization,
corresponding to a supercritical temperature, conventional wisdom states that
the spins organize into a coarsening domain mosaic whose characteristic
length scale grows as $t^{1/2}$~\cite{GSS83,B94,GGGK,book}.  This coarsening
continues as long as the typical domain size is less than the linear
dimension of the system $L$.  At longer times, it is natural to anticipate
that one of the ground states, with magnetization $m=+1$ or $m=-1$, should
ultimately be reached.

For the one-dimensional system, this expectation is fulfilled --- the ground
state is {\em always} reached, independent of the initial condition.  As one
might expect by the diffusive dynamics of the spin domain interfaces, the
characteristic time to reach this final state scales as $L^2$.  In two
dimensions, the ground state is no longer the only asymptotic outcome.  A
system that starts at zero magnetization may also get stuck in an infinitely
long-lived metastable state that consists of straight single-phase
stripes~\cite{SKR01,SKR02,SS,ONSS06}.  The probability to reach such a stripe
state was found numerically to be close to $\frac{1}{3}$.  Recently, a
theoretical argument was given~\cite{BKR09} that relates the stripe state
probability to certain exactly-calculated percolation crossing probabilities.
The resulting probabilities to reach the stripe state are
$\frac{1}{2}-\frac{\sqrt{3}}{2 \pi} \ln \frac{27}{16} = 0.3558\ldots$ for
free boundary conditions and $0.3390\ldots$ for periodic boundary conditions,
in agreement with simulation data~\cite{SKR01,SKR02}.

For regular lattices in three dimensions or greater, little is known about
the evolution and long-time state of this kinetic Ising ferromagnet.  If the
initial magnetization is non-zero, it is generally believed that this system
(on an even-coordinated lattice in $d\geq 2$) ultimately reaches the ground
state of the majority phase in the thermodynamic limit, no matter how small
the initial magnetization~\cite{M10}.  This result has not been proved, even
in two dimensions, although it is plausible because of the connection with
percolation~\cite{BKR09}.  In the $d\to\infty$ limit, the fact that the
(majority) ground state is reached is intuitively obvious and has been
recently proved in Ref.~\cite{M10}.  Physically, however, the zero
initial-magnetization state is much more important than the general case of a
non-zero magnetization.  Indeed, the usual coarsening process begins at a
temperature that exceeds the critical temperature where the initial
magnetization (of an infinite system) equals zero.  We therefore focus on the
case of zero initial magnetization in this paper.

An earlier study~\cite{SKR02} found that an $L\times L\times L$ Ising
ferromagnet on a cubic lattice exhibits a much more complex evolution than
the corresponding two-dimensional system.  In particular, the long-time state
is typically topologically complex and not static.  Here, we investigate the
three-dimensional system in greater detail~\cite{OKR11}.  We offer new
perspectives to help understand the long-time state of the system and present
simulation results to quantify its unusual properties.

\begin{figure}[ht]
\begin{center}
\includegraphics[width=0.275\textwidth]{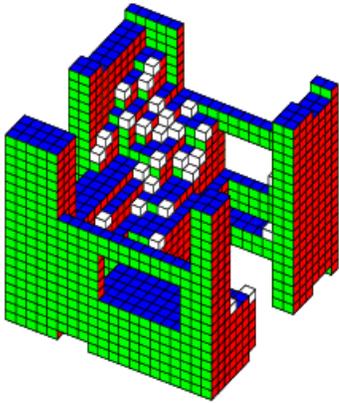}
\caption{\small (color online) A typical example of a state with blinker
  spins for a $20^3$ cubic lattice with periodic boundary conditions.  The
  highlighted blocks indicate blinker spins.}
\label{sponge}
  \end{center}
\end{figure}

\begin{figure}[ht]
\begin{center}
\includegraphics[width=0.145\textwidth]{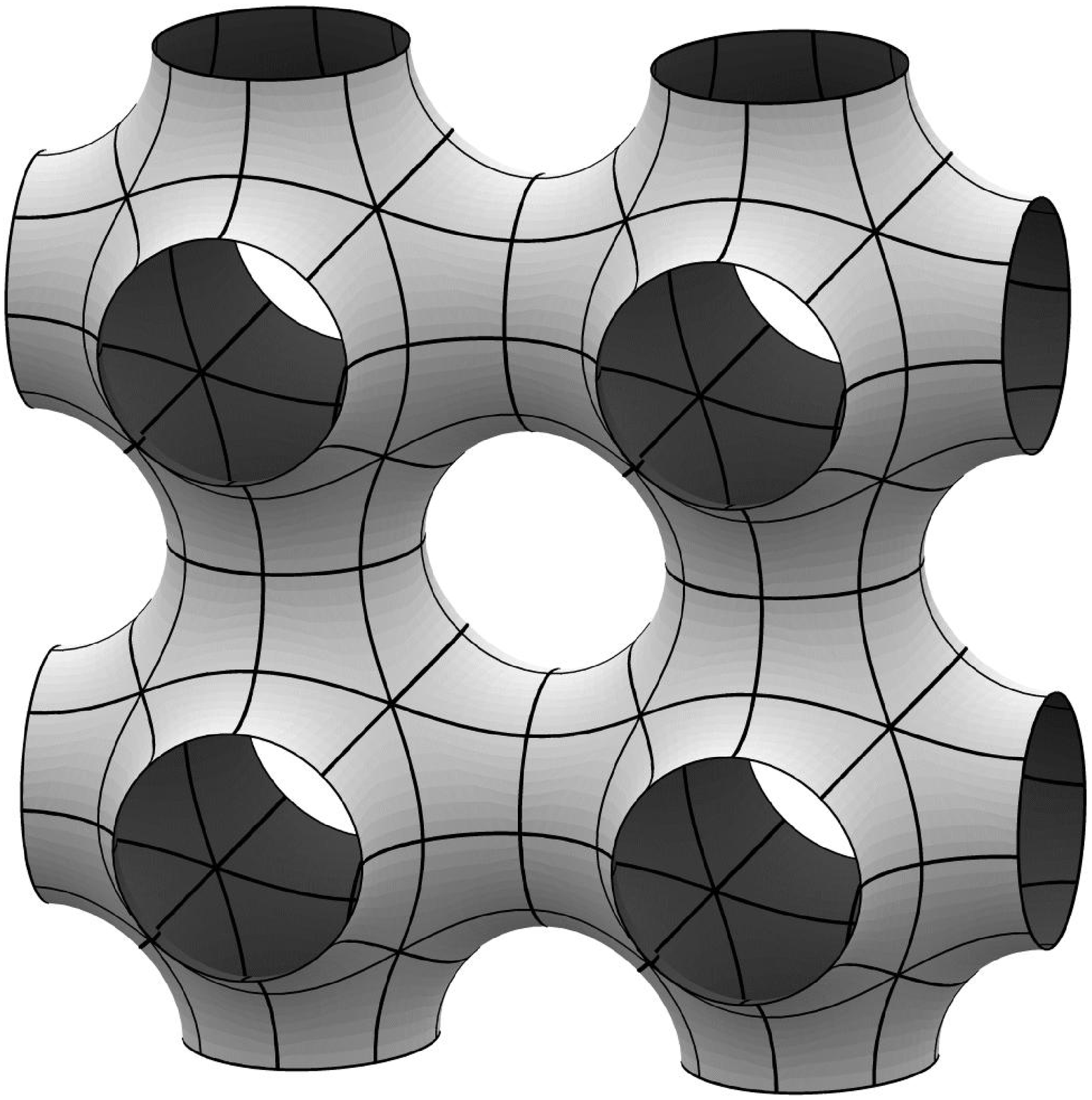}\quad
\includegraphics[width=0.13\textwidth]{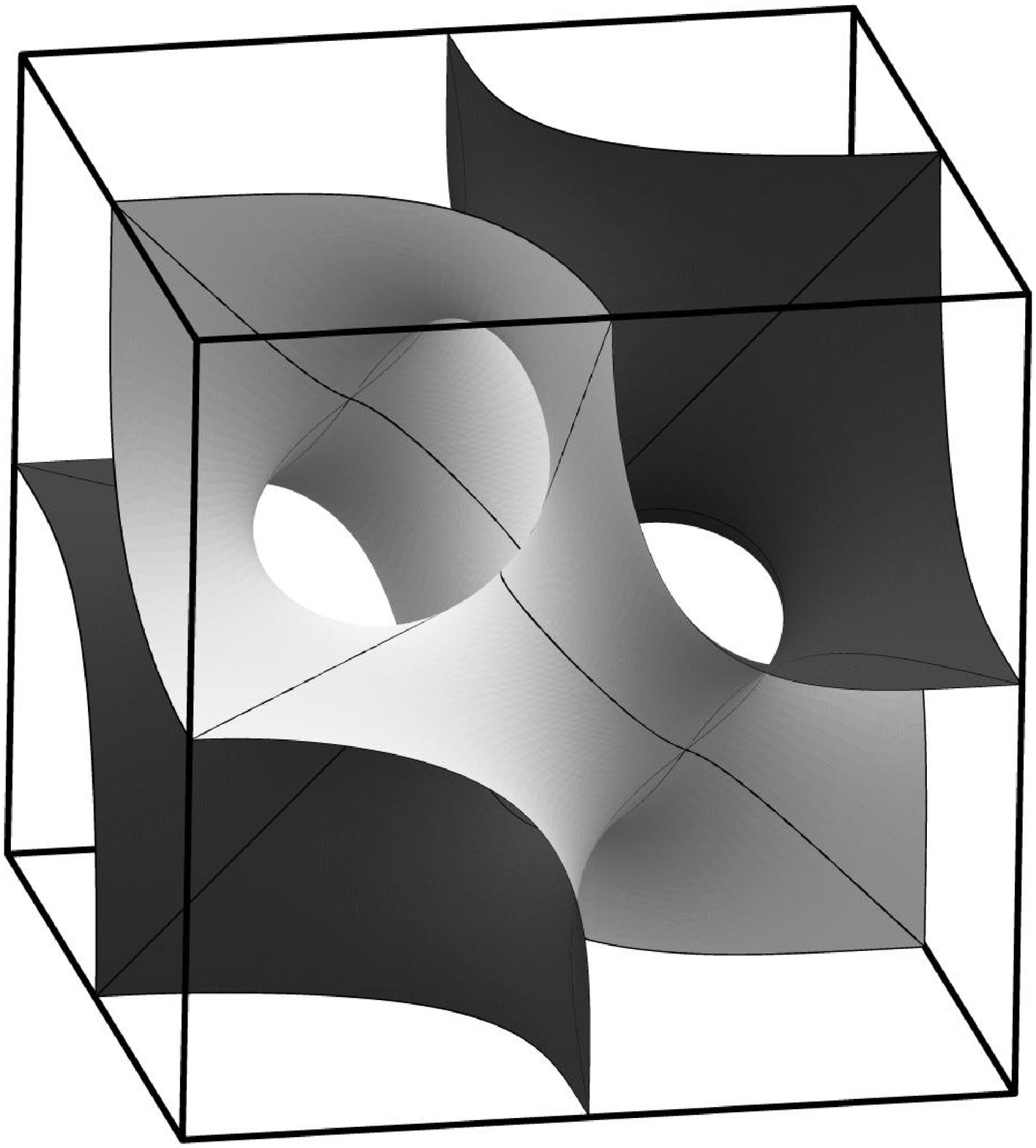}\quad
\includegraphics[width=0.13\textwidth]{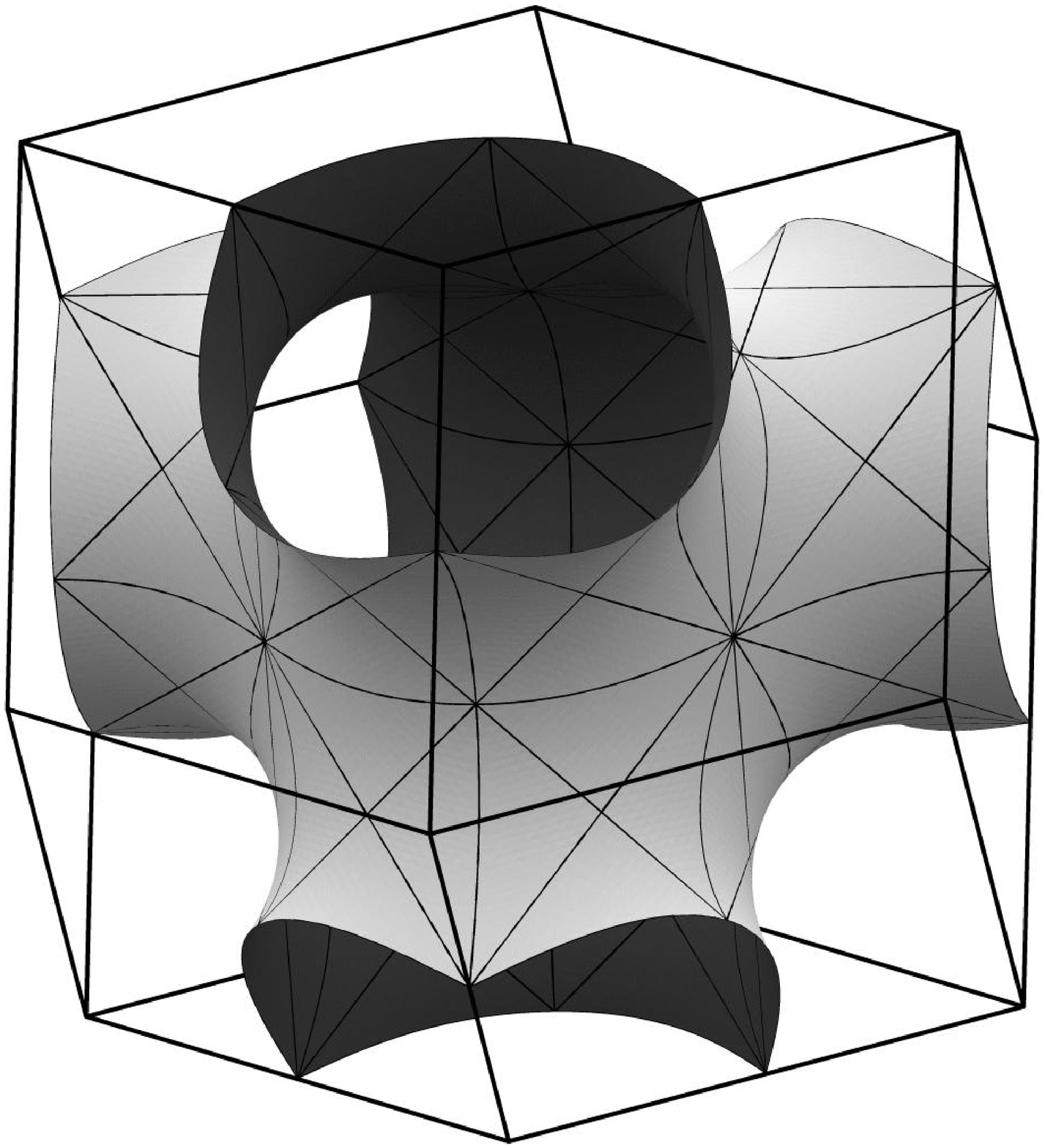}
\caption{\small Triply-periodic minimal surfaces in three dimensions.
  Figures provided by K. Brakke (see also Ref.~\cite{AECTB}).}
\label{periodic}
  \end{center}
\end{figure}

While the ground state is a possible long-time outcome of the dynamics, it
happens that a geometrically rich and infinitely long-lived metastable state
(Fig.~\ref{sponge}) arises with overwhelming probability.  This long-time
state typically has a sponge-like geometry, with multiple interpenetrating
regions of positive and negative magnetization.  The continuum version of
this state is one in which the mean curvature of the interface is zero.  This
restriction leads to a veritable zoo of possible geometries that have been
extensively cataloged~\cite{S90,S69}.  A few illustrative examples of a
subclass of these systems --- periodic zero-curvature continuum interfaces
--- are given in Fig.~\ref{periodic}.  These structures also resemble the
geometrically complex arrangements that arise in two-phase micellar systems,
and are popularly known as ``plumber's
nightmares''~\cite{L89,AS89,FUTGW,AECTB}.

Intriguingly, the long-time state of the three-dimensional system is
generally not static, but rather, contains stochastic {\em blinker} spins
that can flip repeatedly without any energy cost.  The interfaces defined by
these spins can therefore wander {\it ad infinitum} on a connected but
bounded set of equal-energy {\em blinker states} as illustrated in
Fig.~\ref{sponge}.  Here the cubes correspond to up spins (with the spin at
the cube center), while the down spins correspond to blank space.  The
highlighted cubes indicate spins at the convex (outer) corners of domain
interfaces that can flip with no energy cost.  There are also
oppositely-oriented spins (blank spaces) adjacent to the apex of the concave
corners that can also flip up with no energy cost.  A blinker state wanders
perpetually on a small set of iso-energy points in state space.  Blinker
states first appear (albeit rarely) when the linear dimension $L=5$, but
essentially {\em all\/} configurations contain blinkers for large $L$.  While
the fraction of blinker spins is quite small, the fraction of the system
volume over which blinker spins can wander is macroscopic --- it is usually
of the order of ten percent for large $L$.

In Sect.~\ref{model}, we define the system under study.  In Sect.~\ref{sim},
we present a simple method to accelerate the simulations.  Details about the
accuracy of this acceleration algorithm are given in Appendix~\ref{sec:accel}.
We then discuss the physical properties of the long-time state of the system
in Sect.~\ref{long}, including details about blinker states
(Sect.~\ref{blinker}), the asymptotic energy of the system
(Sect.~\ref{energy}), and topological characteristics of the domains
(Sect.~\ref{topology}).  Additional facts about the $L$-dependence of some
basic observables are given in Appendix~\ref{small}.  Typically, domain
interfaces have large genus so that the domains have many interpenetrating
protrusions.  Finally, we investigate the time dependence of the survival
probability (Sect.~\ref{survival}), namely, the probability that the energy
of the system is still decreasing up to a given time.  Based on the insights
gained from studying blinker states, we can understand some important
features of this survival probability.  Concluding remarks are given in
Sect.~\ref{summary}.  Basic features of the evolution of a $2^3$ system,
where all states can be enumerated exactly, as well as a few details of
slightly larger systems are are presented in Appendix~\ref{small}.

\section{Model}
\label{model}

The Hamiltonian of the ferromagnetic Ising model is
\begin{equation}
\label{Ising}
\mathcal{H}=-\sum_{\langle ij\rangle}\sigma_i\sigma_j\,,
\end{equation}
where $\sigma_i=\pm 1$ denotes the spin at site $i$, the interaction strength
is set to one, and the sum is over all nearest-neighbor pairs of sites.  If
not stated explicitly otherwise, we consider the cubic lattice of linear
dimension $L$, with $L$ even, and with periodic boundary conditions.  There
are two natural choices for the initial state: (i) initially uncorrelated and
equal fractions of $+1$ and $-1$ spins (corresponding to an initial
temperature $T=\infty$), or (ii) the antiferromagnetic initial condition, in
which all pairs of neighboring spins are oppositely oriented.  The long-time
evolutions of the system starting from these two initial states are similar,
with only small quantitative differences in the distribution of basic
physical observables, such as the energy and magnetization in the long-time
state.  Thus we focus on the antiferromagnetic initial state for concreteness
and for simplicity.

The spins evolve by zero-temperature single spin-flip Glauber
dynamics~\cite{glauber}.  To implement this dynamics at zero temperature, we
keep a list of flippable spins --- those where the energy change of the
system $\Delta E$ would be zero or negative if the spin were to flip.  (At
zero temperature, spin-flips which would lead to an increase of energy,
$\Delta E>0$, are forbidden \cite{book}.)~ By picking only the spins from
this list, we eliminate the time that would be wasted in picking and
simulating non-flippable spins.  In each update, we pick a flippable spin at
random and flip it with probability 1 if $\Delta E<0$, or with probability
$\tfrac{1}{2}$ if $\Delta E=0$.  This update corresponds to majority rule, as
the condition $\Delta E<0$ means that the majority of neighbors are
antiparallel to the selected spin.

More generally, zero-temperature single spin-flip dynamics is defined by the
rules,
\begin{equation}
  \text{Flipping probability}=
\begin{cases}
1 & \text{if}\quad  \Delta E<0,\\
p & \text{if}\quad \Delta E=0,\\
0 & \text{if}\quad \Delta E>0, 
\end{cases}
\end{equation}
that depend on the single parameter $p$.  For the heat-bath and Glauber
dynamics, $p=\tfrac{1}{2}$, while for the Metropolis algorithm
\cite{Metropolis} all updates proceed with the same rate ($p=1$).  While the
behavior for all $p>0$ is essentially independent of $p$, the $p=0$ case is
special, as the system quickly gets trapped in a jammed configuration
\cite{Luck}.  We shall avoid the $p=0$ dynamics with only strictly
energy-lowering spin flips, as this case is relevant for a special class of
{\em kinetically constrained} systems (see \cite{KC} for review).  Since the
choice for the parameter $p$ is arbitrary (as long as $p$ is strictly
positive), we fix $p=\tfrac{1}{2}$, although this choice is essentially a
matter of habit and, e.g., the Metropolis algorithm with $p=1$ is more
efficient.

For the antiferromagnetic initial condition, all spins are initially
flippable.  The number of flippable spins decreases rapidly at early times
and then the decrease slows down as the system coarsens.  After each update
event, the time is incremented by 1/\#(flippable spins).  Thus in one time
unit, each flippable spin changes its state once on average.  This update
step is applied repeatedly and the dynamics is averaged over many
realizations to determine the evolution of the system.

\section{Acceleration Algorithm}
\label{sim}

\begin{figure}[ht]
\begin{center}
  \includegraphics[width=0.35\textwidth]{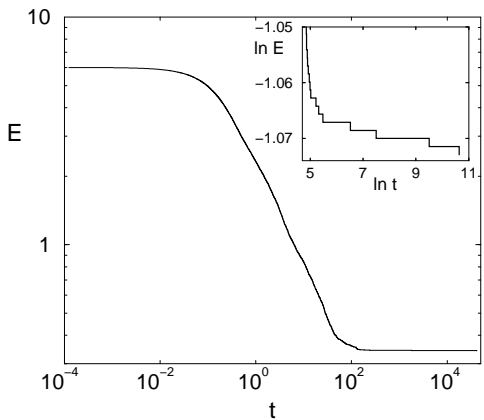}
\smallskip
\caption{\small Time dependence of the energy for a representative
  realization of a $20^3$ system whose energy stops evolving at $t\approx
  41712$.  The inset shows the long-time tail of the same data; note that the
  abscissa is $\ln t$.}
\label{E-single}
  \end{center}
\end{figure}

The evolution of the system becomes so slow that the standard Glauber
dynamics algorithm described above is inadequate to probe the long-time
properties of even reasonably-sized systems.  Figure~\ref{E-single}
illustrates this slowing down for the energy evolution of a typical
realization of a $20^3$ system.  The main panel shows the time dependence of
the gap between the actual energy and the ground state energy divided by the
total number of spins.  Henceforth, we term this normalized difference as the
``energy'' $E_L$.  For this example, the data are roughly consistent with the
$E_L\sim t^{-1/2}$ for $1\alt t\alt 10^2$.

\begin{figure}[ht]
\begin{center}
\includegraphics[width=0.35\textwidth,height=0.08\textwidth]{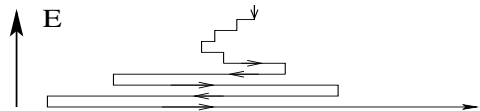}
\smallskip
\caption{\small Schematic illustration of the state of the system wandering
  on  fixed-energy plateaux at long times. }
\label{plateau}
  \end{center}
\end{figure}

\begin{figure}[ht]
\begin{center}
\includegraphics[width=0.35\textwidth]{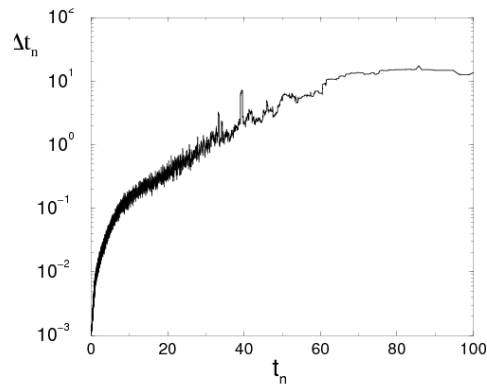}
\smallskip
\caption{\small Average time $\Delta t_n$ between successive energy-lowering
  spin-flip events as a function of $t_n$, the time for the $n^{\rm th}$ such
  event for $1024$ realizations of a $10\times 10\times 10$ system.  The data
  are smoothed over a 100-point range. }
\label{t-vs-n}
  \end{center}
\end{figure}

For times beyond the coarsening time (which scales as $L^2$), the energy
evolution is characterized by long periods where only zero-energy spins can
flip (those with equal numbers of up and down neighbors).  These long periods
of stasis are punctuated by progressively more rare energy-decreasing events
(inset to Fig.~\ref{E-single}).  Thus the system typically wanders on
successive plateaux that define a set of iso-energy points in state space.
Occasionally there is a drop to a lower plateau by an energy-lowering
spin-flip event (Fig.~\ref{plateau}).  This feature is illustrated in
Fig.~\ref{t-vs-n}, where we plot the configuration averaged time $\Delta t_n$
between successive energy-lowering spin-flip events as a function of $t_n$,
the average time at which the $n^{\rm th}$ such event occurred.  Over a
substantial range, $\Delta t_n$ appears to grow exponentially with $t_n$, so
that most of the evolution is spent wandering aimlessly on iso-energy
plateaux.  A somewhat related continuum picture of this state space evolution
is presented in Refs.~\cite{KL96,F97}.

To reduce the time spent in simulating these iso-energy wanderings, we
constructed an acceleration algorithm and tested that the long-time state
achieved by this algorithm is virtually identical to that of the true
zero-temperature Glauber dynamics.  Our method is based on imposing a weak
field while the system is on any single iso-energy plateau to reduce the time
needed to find the next energy lowering event.  The field is reversed after
each such event so that the average field is zero.  The steps of our method
are the following:
\begin{itemize}
\itemsep -0.5ex
\item Apply Glauber dynamics (to flippable spins only) until $t=5L^2$.  This
  time is sufficiently beyond the coarsening time that energy-lowering events
  have become rare (Fig.~\ref{E-single}).

\item For $t>5L^2$, apply an infinitesimal field to drive the state-space
  motion along a fixed-energy plateau.  Thus the next energy-lowering event
  (if it exists) is found more rapidly than if no field was applied.

\item After each energy-lowering event occurs, the sign of the field is
  reversed, leading to the alternating state-space motion sketched in
  Fig.~\ref{plateau}.

\item If the number of active spins goes to zero without a drop in energy
  while the field is applied, the field is reversed.  From this
  configuration, the system evolves by zero-temperature Glauber dynamics with
  the reversed field.  If the number of active spins again goes to zero
  without a drop in energy, then the final energy value has been reached and
  the simulation is finished.
\end{itemize}

We verified that this acceleration algorithm accurately reproduces the energy
obtained by straightforward Glauber dynamics for $L\leq 10$, where a direct
check of this acceleration method is computationally feasible.  For this
check, we take all configurations that have been evolved to $t=5 L^2$ by
zero-temperature Glauber dynamics and evolve each one both by continuing the
Glauber dynamics and by our acceleration algorithm until no flippable spins
remain.  Over $10^7$ realizations the fractional difference between the
energies obtained by these two methods is $\leq 1.4\times 10^{-8}$.
Moreover, the distributions of the final energies are virtually
indistinguishable.

For larger $L$, time to reach the final energy by zero-temperature Glauber
dynamics is too long to amass sufficient statistics.  Instead, we compare the
ultimate energy that is reached by starting the acceleration algorithm at
progressively later times.  As shown in the Table~\ref{tab:compare} in
Appendix~\ref{sec:accel}, the energy that is ultimately reached changes by
a negligible amount as the cutoff time is increased.  For example, for
$L=100$, the error in the final energy reached by the acceleration algorithm
is of the order of $5\times 10^{-5}$.  Moreover, as illustrated in this
table, the acceleration algorithm is considerably faster than the
zero-temperature Glauber dynamics.  Thus we use the acceleration algorithm
for all of our simulation results.  On a 32-core machine, we are able to
simulate $10^5$ realizations of a $90^3$ system in approximately ten days of
running time.

\section{The Long-Time State}
\label{long}

\begin{figure}[ht]
  \includegraphics[width=0.35\textwidth]{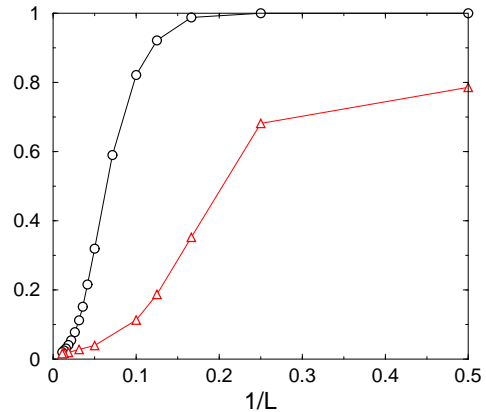}
  \caption{\small Plot of $1-P_b$, the complement of the probability $P_b$ of
    reaching a blinker configuration ($\circ$), and $1-P_2$, the complement of
    the probability $P_2$ of reaching a state that consists of 2 clusters
    ($\bigtriangleup$), as a function of $1/L$.}
\label{blinker-prob}
\end{figure}

At sufficiently long times, the energy of each realization stops decreasing,
either because a blinker configuration is reached or occasionally a static
final state is reached.  Figure~\ref{blinker-prob} shows that the probability
of reaching a blinker configuration approaches 1 as $L\to\infty$.  This
observation is one of our main results, for which preliminary corroboration
was given earlier~\cite{SKR01,SKR02}.  Somewhat surprisingly, for the initial
condition where the magnetization is zero but with otherwise uncorrelated
spins, the probability to reach a stationary state vanishes more rapidly in
the $L\to\infty$ limit than for the antiferromagnetic initial condition.
Even though a blinker consists of a set of states of the same energy, we term
the ``final state'' these constant-energy configuration(s) that are reached
when the energy stops decreasing.

In this final state, the spins almost always partition into two and only two
interpenetrating clusters.  Using the Hoshen-Kopelman cluster-multilabeling
algorithm~\cite{HK76} to determine the distribution of the number of
clusters, we find that the probability to find a single cluster
(corresponding to the ground state) or more than two clusters in the final
state rapidly decays with $L$ (Fig.~\ref{blinker-prob}).  Final states that
contain more than two clusters typically consist of multiple narrow filaments
of one phase in a surrounding background of the opposite phase.  In all of
our simulations on the $L\times L\times L$ cubes with periodic boundary
conditions, the largest number of clusters observed in any realization was
seven.  (This happened to occur in a $38^3$ system, where the realization
consisted of six narrow filaments of one phase in a background of the
opposite phase.)

\subsection{Relaxation of Blinker Configurations}
\label{blinker}

\begin{figure}[ht]
\begin{center}
\includegraphics[width=0.14\textwidth]{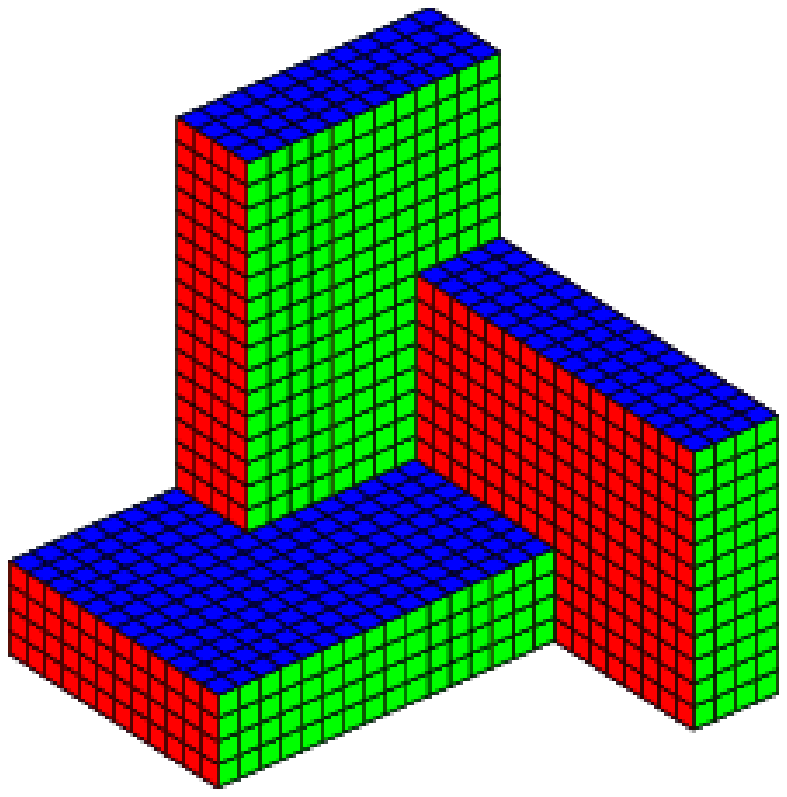}\quad
\includegraphics[width=0.14\textwidth]{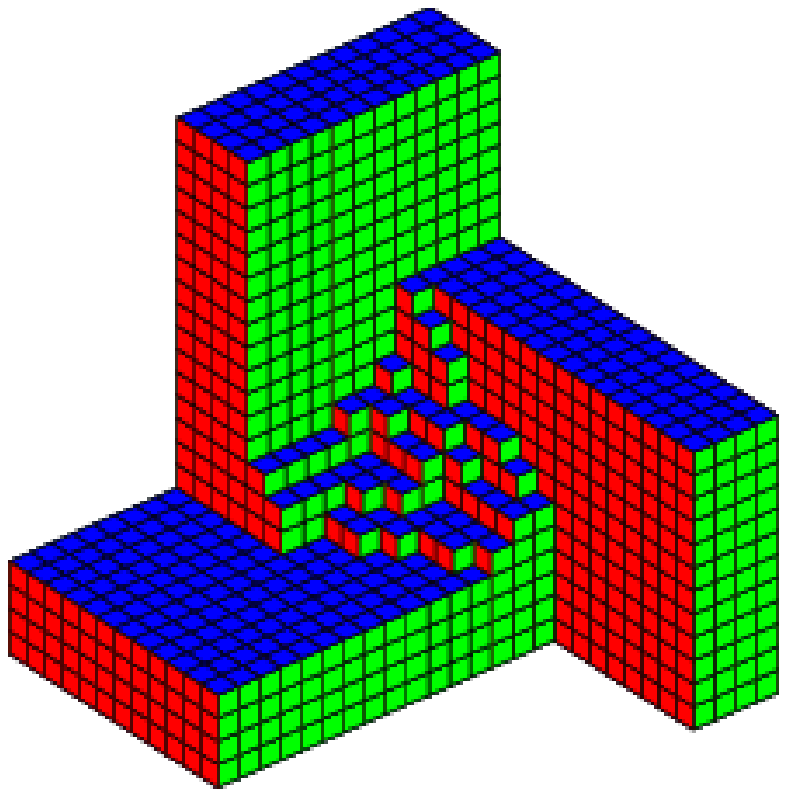}\quad
\includegraphics[width=0.14\textwidth]{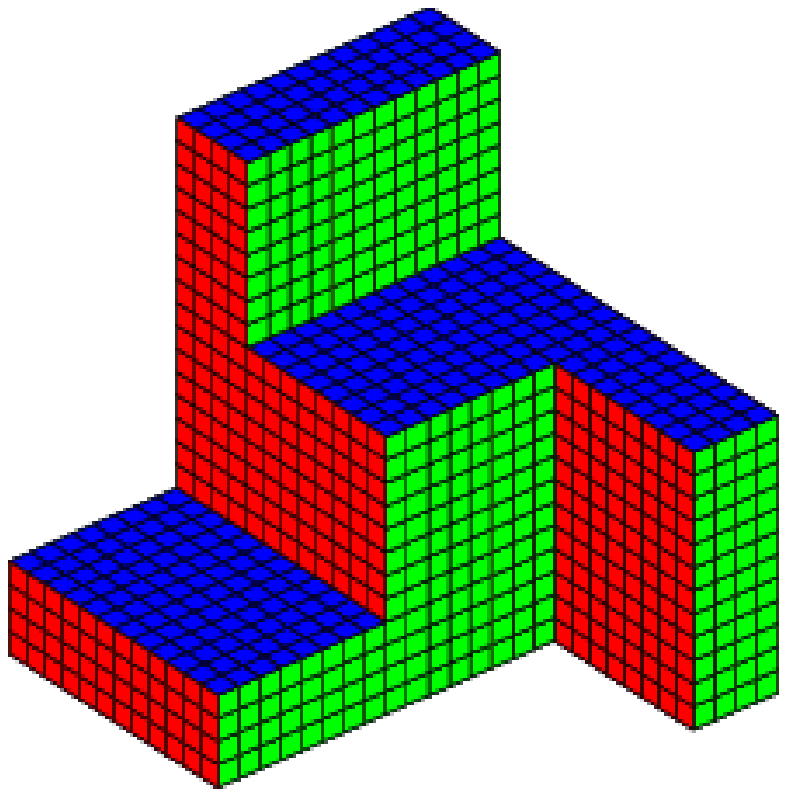}
\smallskip
\caption{\small (color online) An $8^3$ blinker on a $20^3$ cubic lattice,
  showing the fully-deflated state (left), an intermediate state (middle),
  and the fully-inflated state (right).  The bounding slabs wrap periodically
  in the three Cartesian directions.}
\label{blinker-evol}
  \end{center}
\end{figure}

As we now discuss, blinker states are responsible for the very slow
relaxation of the spin system at long times.  To appreciate the underlying
mechanism, it is instructive to study the dynamics of the synthetic blinker
states shown in Fig.~\ref{blinker-evol}.  In this example, the domain of up
spins consists of three orthogonal $4\times 12$ slabs, each of which wraps
periodically, so that the apparent slab corners are merely visual artifacts.
In the cavity defined by the confluence of the three slabs, a blinker state
exists.  By zero-energy spin flips, the portion of the interface defined by
this cavity can be in the extremes of fully deflated (left panel of
Fig.~\ref{blinker-evol}), or fully inflated (right), or in some intermediate
state (middle).  Note that each extreme configuration possesses a single
blinker spin, while intermediate configurations have more than one blinker
spin.  Although each blinker spin does not experience any energetic bias,
there is an effective geometric bias that drives the interface to the
half-inflated state.  This effective bias is controlled by the difference in
the number of flippable spins on the convex and concave corners on the
interface, $N_+$ and $N_-$, respectively.  When the cube is mostly inflated
$N_+-N_-$ is positive so that the interface tends to deflate, and vice versa
when the cube is mostly deflated.  Thus the effective bias drives the
interface to the half-inflated state.

\begin{figure}[ht]
\begin{center}
\includegraphics[width=0.13\textwidth]{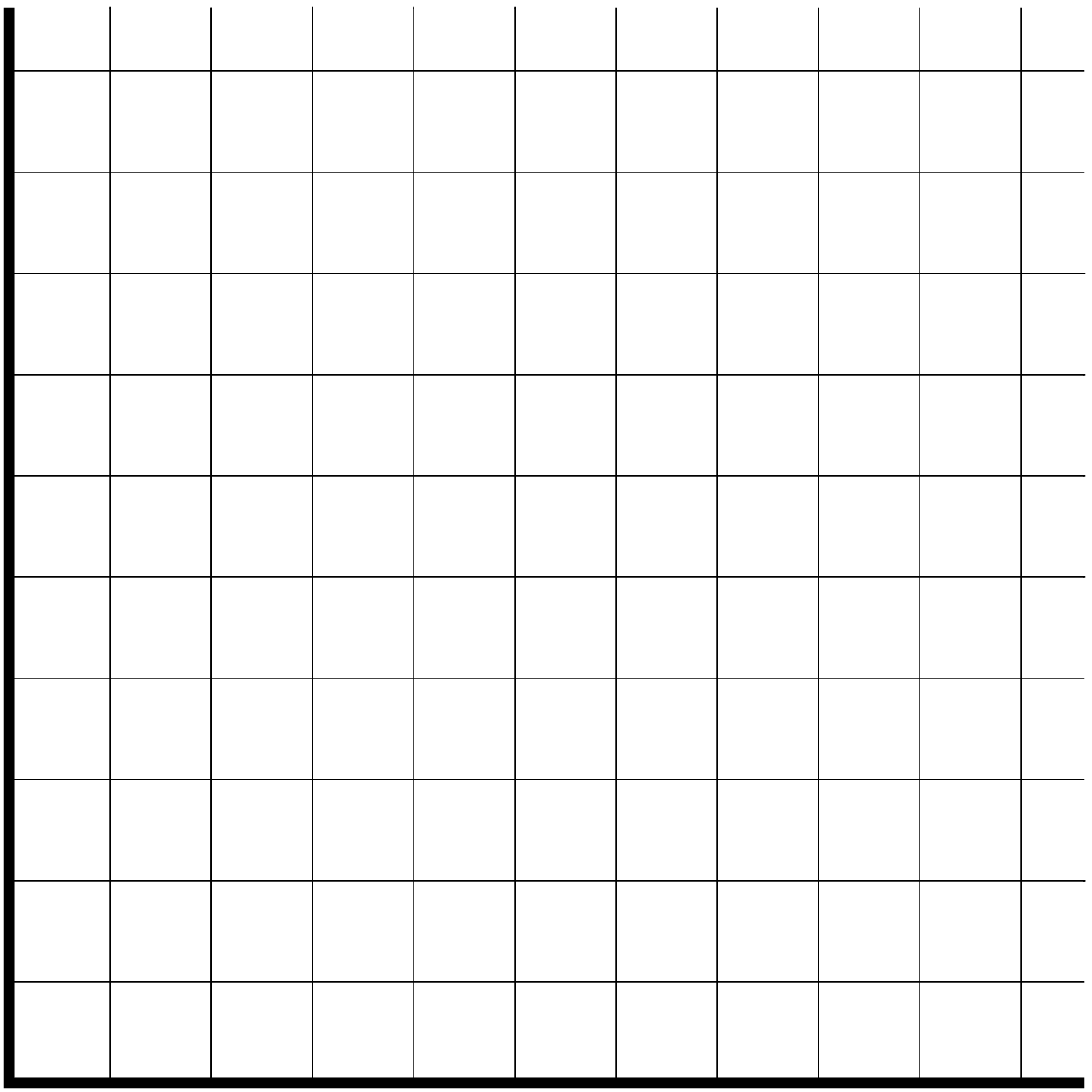}\qquad
\includegraphics[width=0.13\textwidth]{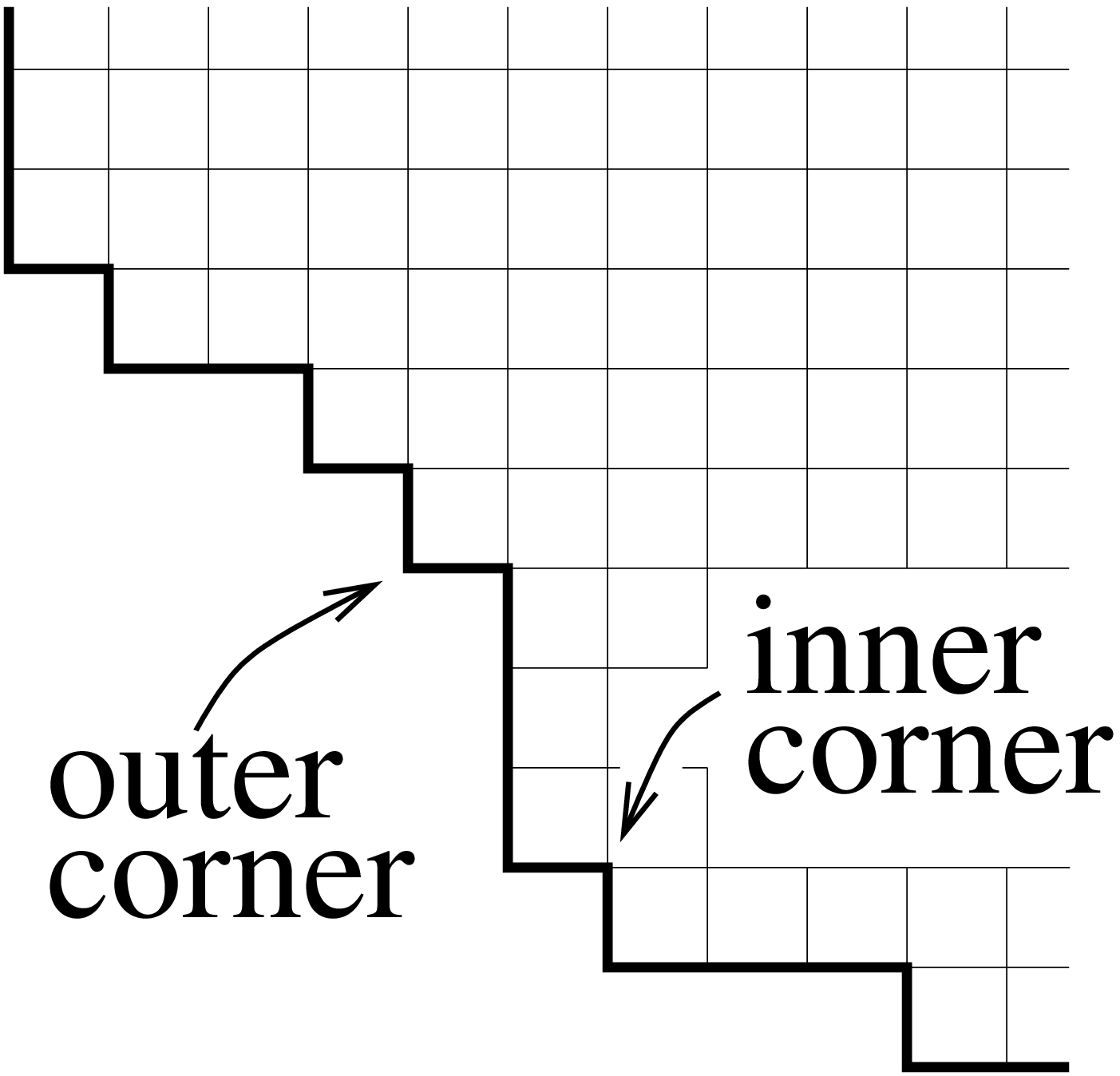}\qquad
\includegraphics[width=0.13\textwidth]{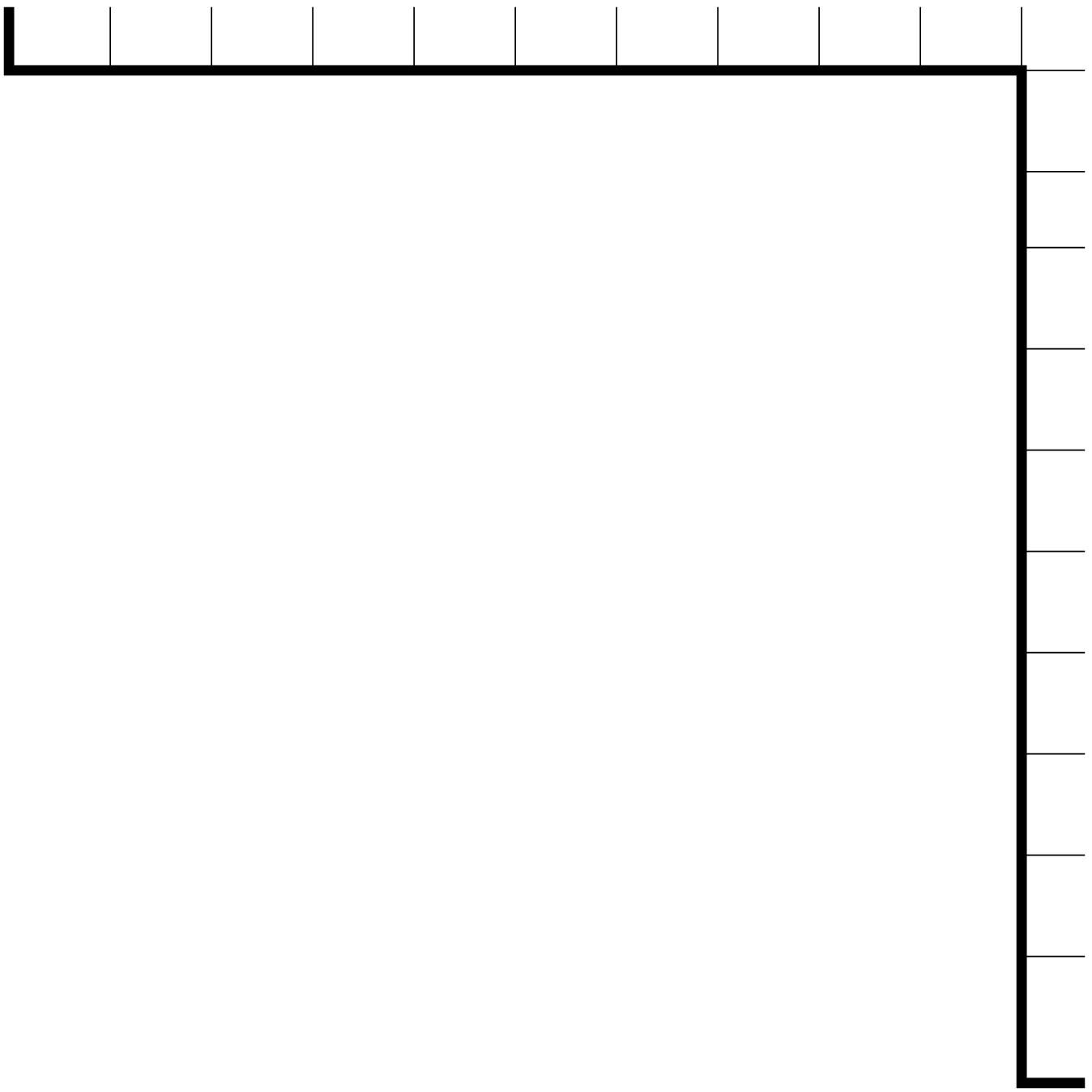}
\smallskip
\caption{\small Two-dimensional analog of the blinker states in
  Fig.~\ref{blinker-evol}. }
\label{blinker-2d}
  \end{center}
\end{figure}

We quantify the evolution of this synthetic blinker by the average
first-passage time $\langle t\rangle$ for an $\ell\times \ell\times \ell$
fully-deflated blinker to become fully inflated.  To estimate this time, it
is helpful to first consider the corresponding two-dimensional system
(Fig.~\ref{blinker-2d}).  Near the half-inflated state, the interface
consists of $N_+$ outer corners and $N_-$ inner corners, with $N_+-N_-=1$ in
two dimensions and $N_+\sim \ell$.  In a single time unit, all eligible spins
on the interface flip once, on average.  Since $N_+-N_-=1$, the interface
area typically decreases by 1.  Thus we infer an interface velocity $u=\Delta
A/\Delta t\sim -1$.  Similarly, the mean-square change in the interface area
is of the order of $N_+\sim\ell\sim \sqrt{A}$.  Thus the effective diffusion
coefficient $D$ is proportional to $\ell$.  The first-passage time is
dominated by the time to move from the half-inflated state to the
fully-inflated state by flipping of the order of $A=\ell^2$ spins.  Since
this process is moving against the effective bias, the dominant Arrhenius
factor~\cite{fpp} in the first-passage time is $\tau\sim \exp(|u|
\ell^2/D)$, so that
\begin{equation}
\label{T2}
\ln \tau\sim  \ell\,.
\end{equation}

For the corresponding three-dimensional system of volume $V=\ell^3$, there
are typically $N_\pm\sim \ell^2$ outer and inner corners when the interface
is half inflated.  The disparity in their number is now of the order of
$\ell$.  Thus in a single time step the displacement of the interface is
of the order of $\ell$; this quantity coincides with the interface velocity.
Similarly, the mean-square change in the interface volume in a unit time,
which coincides with the diffusivity, is of the order of $N_\pm\sim
\ell^2\sim D$.  Consequently, the leading behavior of the first-passage time
is
\begin{equation}
\label{T3}
\ln \tau\sim u \ell^3/D\sim \ell^2\,.
\end{equation}
The direct generalization of this argument to $d$ dimensions gives
$\ln\tau\sim \ell^{d-1}$.  Related aspects of slow evolution in three
dimensions were discussed for the homogeneous kinetic Ising
ferromagnet~\cite{L99} and for a kinetic Ising system with competing
ferromagnetic and antiferromagnetic interactions~\cite{SHS}.

\begin{figure}[ht]
\begin{center}
\includegraphics[width=0.35\textwidth]{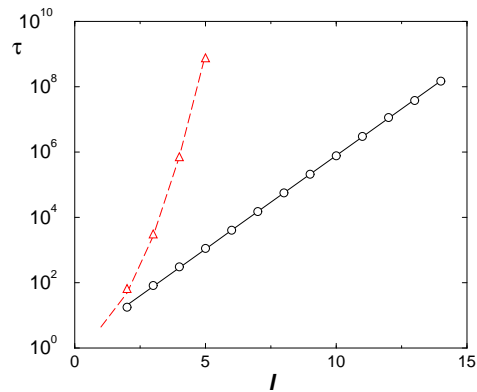}
\smallskip
\caption{\small First-passage time from the fully-deflated to the
  fully-inflated state for $d=2$ ($\circ$) and $d=3$ ($\bigtriangleup$).  The
  line for $d=2$ is the best fit $\tau=1.40\,\exp(1.33\,\ell)$, while the
  curve for $d=3$, $\tau= 3\,\exp(0.8\,\ell^2)$, is merely a guide for the
  eye.  Each data point is based on at least 128 realizations.}
\label{t-blinker}
  \end{center}
\end{figure}

Figure~\ref{t-blinker} shows simulation data for
the first-passage time from the fully-deflated to the fully-inflated state in
two and three dimensions.  The agreement between Eq.~\eqref{T2} and the
two-dimensional data is excellent.  In three dimensions, simulations are
necessarily limited to very small $\ell$, while our crude argument is
asymptotic; nevertheless, the data are qualitatively consistent with
Eq.~\eqref{T3}.  The salient result is that the time for a fully-deflated
blinker to become fully inflated grows rapidly with $\ell$.

\begin{figure}[ht]
\includegraphics[width=0.325\textwidth]{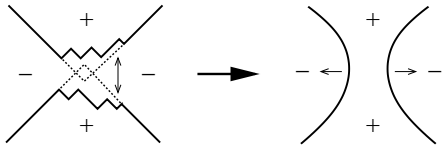}
\caption{A schematic two-dimensional projection of a blinker coalescence
  event.  The tips of the blinkers correspond to the fully-inflated
  configuration shown in the right panel of Fig.~\ref{blinker-evol}.}
\label{blinker-merge}
\end{figure}

From the dynamics of the synthetic blinker of Fig.~\ref{blinker-evol}, we can
now understand the long time scales associated with the relaxation of a large
system.  Indeed, consider two such blinkers that are oppositely oriented and
spatially separated so that they do not overlap when both are half inflated,
but just touch corner to corner when both are inflated
(Fig.~\ref{blinker-merge}).  As long as the blinkers do not overlap, their
fluctuations do not change the energy of the system.  However, if these
blinkers touch, then a spin flip event has occurred that lowers the energy.
After this irreversible coalescence, subsequent spin flip events cause the
two blinkers to ultimately merge.

Because of the impermanence of a two-blinker configuration, we term it a {\em
  pseudo-blinker}.  We assert that each coalescence of a pseudo-blinker
corresponds to one of the increasingly rare energy-lowering spin-flip events
sketched in Fig.~\ref{plateau}.  The time for pseudo-blinker coalescence is
extraordinarily long because the time for each blinker to reach a
nearly-inflated state is a rapidly increasing function of its size. These
coalescences correspond to energy-lowering spin-flip events at long times.

We now turn to the true blinker states.  As might be
anticipated from the example in Fig.~\ref{sponge}, simulations indicate that
the fraction of blinker spins is small --- of the order $3\times 10^{-3}$ to
$4\times 10^{-3}$ of all spins for systems of linear dimension $ L \leq 50$.
The instantaneous number of blinker spins also fluctuates substantially so
that their number is not a meaningful characteristic of the blinker states.
A more robust measure is the total volume that is accessed by blinker spins.
We determine this accessible volume as follows: Once a true blinker state is
first reached (which we define as $\mathcal{B}_0$), we drive the system with
an infinitesimal positive field until the spin configuration $\mathcal{B}_+$,
with no flippable spins, is reached.  Then starting again from
$\mathcal{B}_0$, we drive the system with an infinitesimal negative field
until there are no flippable spins and the configuration $\mathcal{B}_-$ is
reached.  The difference $|\mathcal{B}_+-\mathcal{B}_-|$ defines the total
blinker volume.  The resulting blinker volume fraction is a slowly increasing
function of $L$ and extrapolating to $L\to\infty$ gives an asymptotic blinker
volume fraction of approximately 9\% --- a finite fraction of the entire
system.

\subsection{Asymptotic Energy}
\label{energy}

\begin{figure}[ht]
\includegraphics[width=0.35\textwidth]{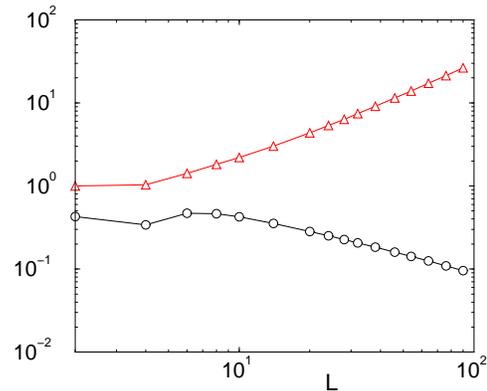}
\caption{\small Normalized average energy $(\circ)$ and genus of the final
  state ($\bigtriangleup$) as a function of $L$.  The relative error for each
  data point is less than $1.4\times 10^{-3}$.}
\label{fig:Eg}
\end{figure}

An important characteristic of a finite system of linear dimension $L$ is its
energy $E_L$ at infinite time.  As mentioned in Sect.~\ref{sim}, what we term
the energy is actually the energy gap above the ground state energy per spin.
This energy decreases with $L$ in a manner consistent with the power-law
dependence ${E}_L\sim L^{-\epsilon}$ (Fig.~\ref{fig:Eg}).  However, there is
systematic curvature in this data, and we extrapolate the local two-point
slopes in the plot of $E_L$ versus $L$ to obtain the estimate $\epsilon
\approx 1$, in agreement with previous results based on smaller-scale
simulations \cite{SKR02}.  This dependence implies that the total interface
area between spin domains scales as $L^2$.

\begin{figure}[h]
\includegraphics[width=0.35\textwidth]{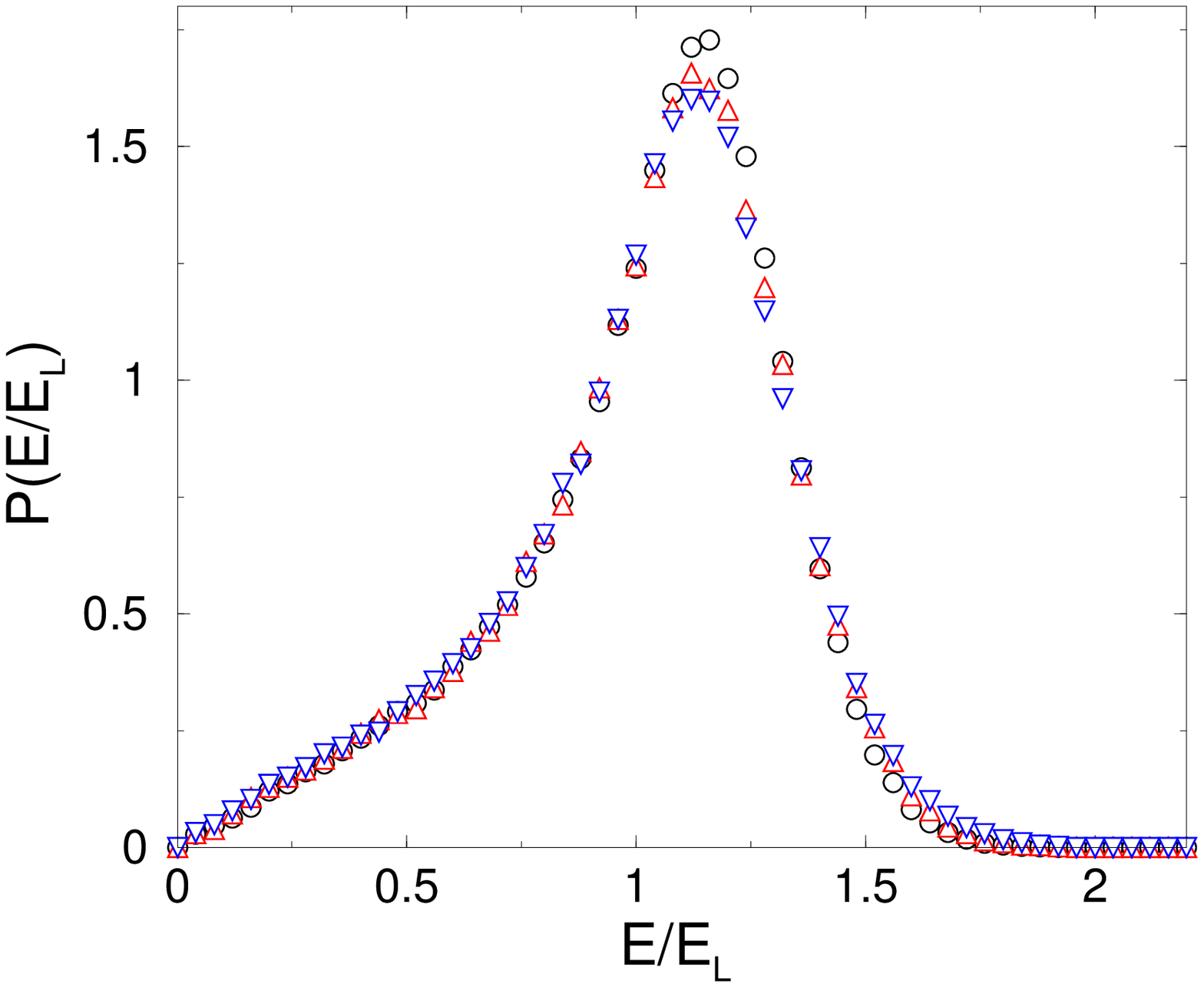}
\caption{\small The normalized final-state energy distribution for $L=54$
  ($\circ$), $L=76$, ($\bigtriangleup$), and $L=90$ ($\bigtriangledown$).}
\label{fig:energy-dist}
\end{figure}

The scaled distribution of energies, $P(E/E_L)$, exhibits an 
excellent data collapse (Fig.~\ref{fig:energy-dist}).  The distribution has a
well-defined peak that is close to a Gaussian; there is also a noticeable
linear tail at low energies. The fact that the energy distribution 
$P(E/E_L)$ remains broad in the thermodynamic limit is not 
surprising as all our numerical findings show the lack of self-averaging. 
It would be interesting to understand qualitatively the shape of the energy distribution 
$P(E/E_L)$, or at least the asymptotic
behaviors in the $E/E_L\to 0$ and $E/E_L\to \infty$ limits.

\subsection{Domain Topology}
\label{topology}

The long-time state of the three-dimensional system is, in general,
topologically complex because it is possible --- in fact likely --- that the
interface contains many holes~\cite{2d}.  The number of holes is also known
as the genus $g$.  Formally, the genus of a connected, orientable surface is
an integer that represents the maximum number of cuts that can be made
through the surface along non-intersecting closed simple curves without
disconnecting the resulting manifold.  As elementary examples, the genus of a
sphere is $g=0$, while the genus of a doughnut is $g=1$.

\begin{figure}[ht]
  \includegraphics[width=0.1\textwidth]{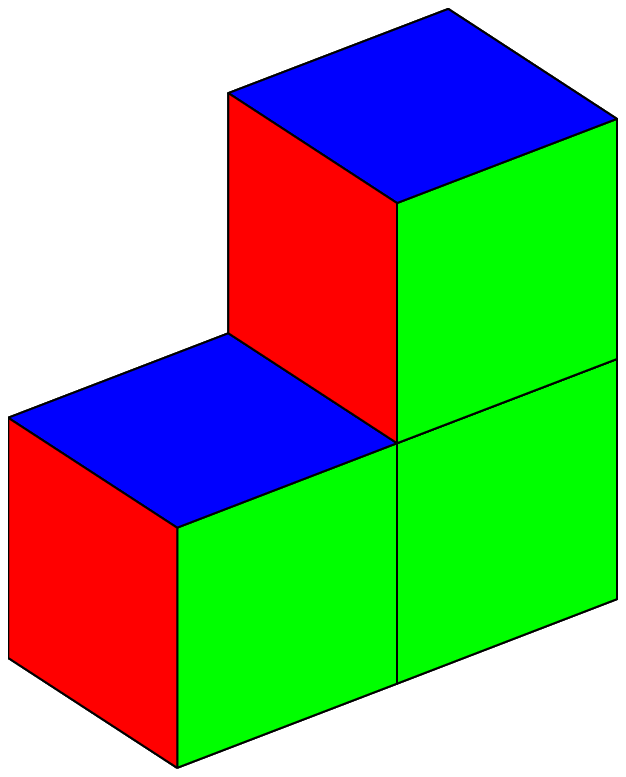}\hskip
  0.5in\includegraphics[width=0.125\textwidth]{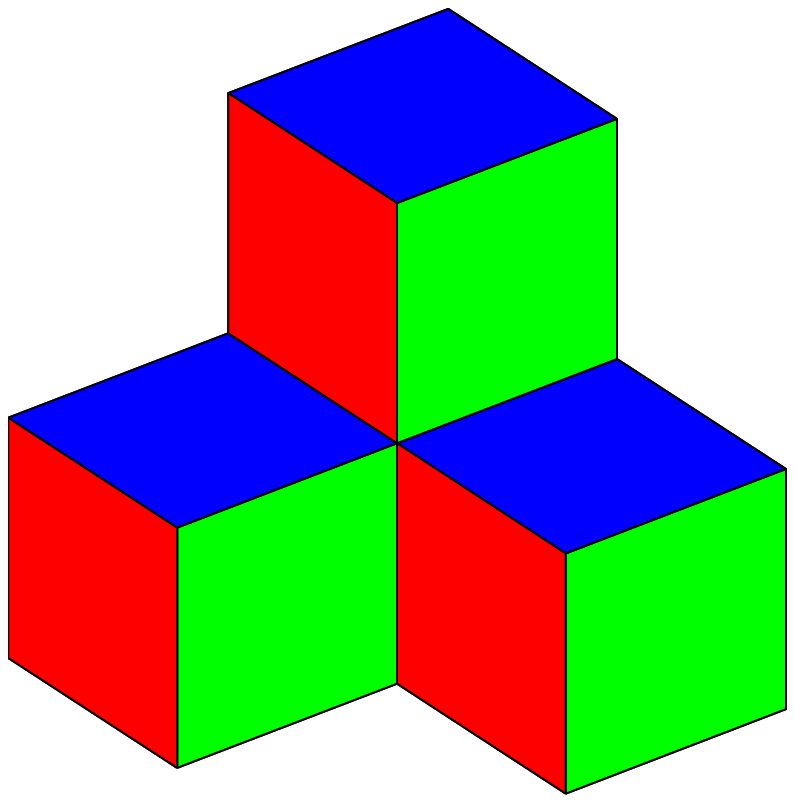}
\caption{(color online) Simple examples of interfaces with genus $g=2$ and
  $g=3$ for a periodic $2\times 2\times 2$ system.}
\label{fig:g}
\end{figure}

\begin{figure}[ht]
\includegraphics[width=0.2\textwidth]{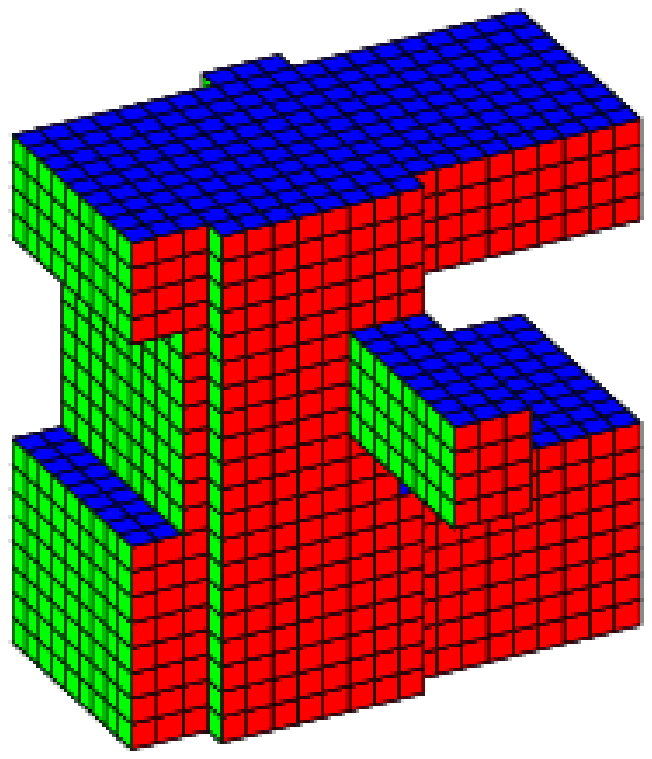}\quad
\includegraphics[width=0.2\textwidth]{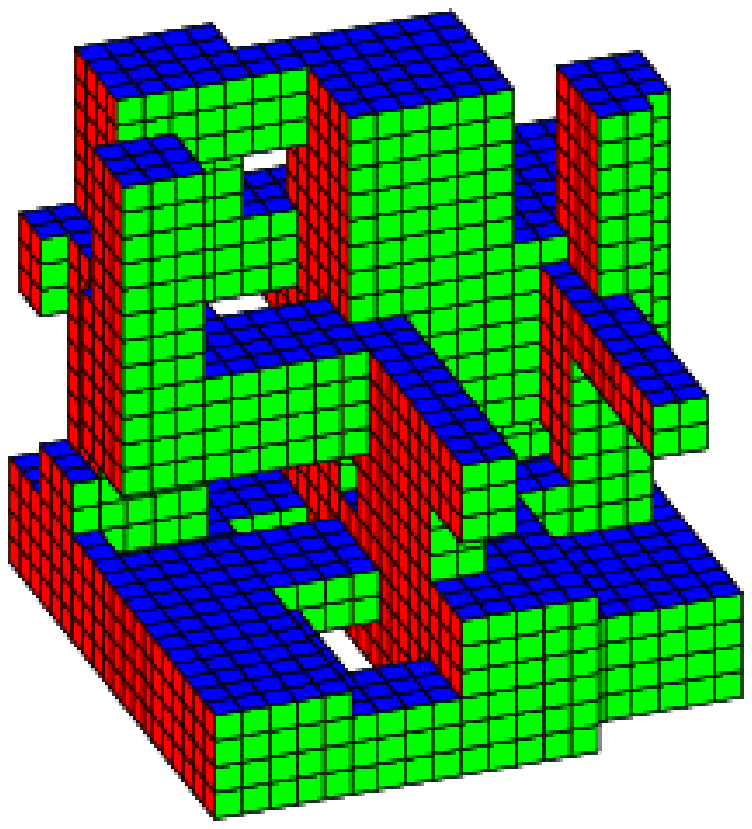}
\caption{(color online) Examples of a low-genus ($g=3$) and high-genus
  ($g=16$) domain on a $20^3$ lattice.}
\label{fig:g-20}
\end{figure}

The genus of a surface can be expressed through its Euler characteristic $\chi$ 
(see e.g. \cite{E,Japan})
\begin{equation}
\label{Euler}
\chi= 2(1-g)= \mathcal{V}-\mathcal{E}+\mathcal{F}.
\end{equation}
The latter equality relates the Euler characteristic to easily-measured
features of the domain interface: $\mathcal{V}$, the number of vertices on
the interface, $\mathcal{E}$, the number of edges, and $\mathcal{F}$, the
number of faces.  As simple examples, the Euler characteristic of an isolated
cube is $\chi=8-12+6=2$, corresponding to genus 0.  (Topologically, the cube
is identical to the ball, so the boundary of the cube is a sphere.)~ The
Euler characteristic of a linear filament that wraps around the torus in one
direction is zero.  By discretizing this filament as a $2\times 1$ cluster
that wraps onto itself, we have $\mathcal{V}=8$, $\mathcal{E}=16$ and
$\mathcal{F}=8$, corresponding to genus $g=1$.  Note that the Euler
characteristic does not depend on the length scale of the discretization.
Similarly for a cluster that percolates in two directions (Fig.~\ref{fig:g}),
$\chi=8-20+10=-2$ so the genus $g=2$.  Finally for a cluster that percolates
in all three Cartesian directions, $\chi=8-24+12=-4$ so the genus $g=3$.

Since blinker spins do not affect the topology, we freeze these spins in
their orientations at the time when the interface topology is measured.  To
measure the topology, we first identify all the clusters in the final state
by the Hoshen-Kopelman algorithm~\cite{HK76}.  We then compute the Euler
characteristic of each cluster.  If there are only two clusters, then by
construction they have the same interface and thus the same Euler
characteristic.  If a final state has more than two clusters, we use the
maximum genus among all clusters as the genus of the system.  To determine
the Euler characteristic numerically, we first determine the number of faces
$\mathcal{F}$.  This quantity equals the number of neighboring antiparallel spins
(accounting for the periodic boundary conditions) which, in turn, is directly
related to the energy of the system.  Once we identify a new face on the
interface, each of the four vertices and the four edges bounding this face
are added to the current counts of $\mathcal{V}$ and $\mathcal{E}$, as long
as they have not yet already been counted as part of another
previously-encountered face.

\begin{figure}[h]
\includegraphics[width=0.35\textwidth]{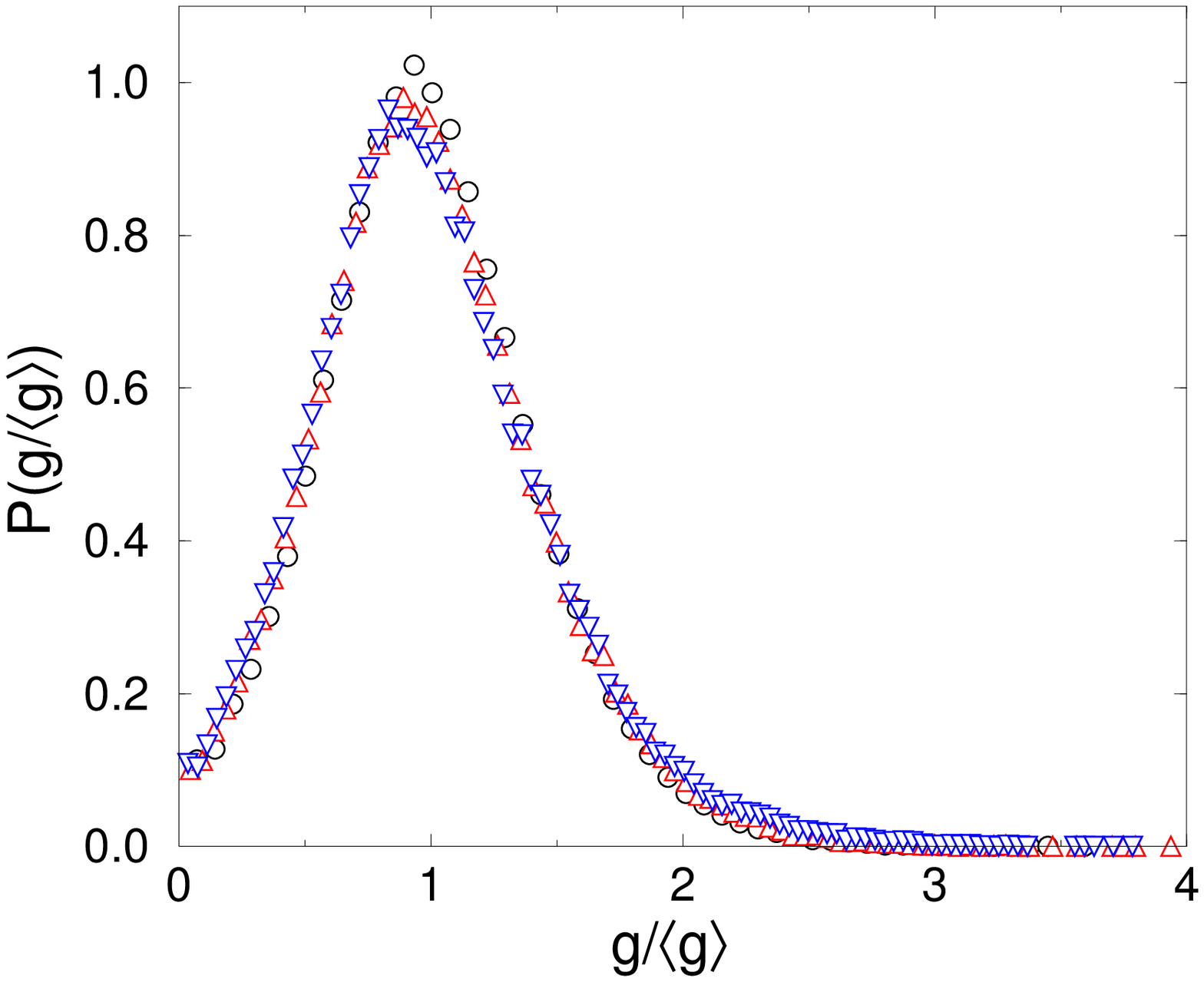}
\caption{\small The final-state genus distribution for $L=54$ ($\circ$),
  $L=76$, ($\bigtriangleup$), and $L=90$ ($\bigtriangledown$).}
\label{fig:genus-dist}
\end{figure}

The resulting domain topologies are quite diverse (Fig.~\ref{fig:g-20}).  For
example, for $L=20$, the smallest genus observed was 1, the largest genus was
18, while the average genus is approximately 4.36.  The average genus for a
given $L$, defined as $g_L$, again appears to grow as a power law in $L$,
$g_L\sim L^\gamma$, but with substantial finite-size corrections
(Fig.~\ref{fig:Eg}).  Analogous to the behavior for the average energy, the
data for $g_L$ versus $L$ on a double logarithmic scale are systematically
curved upward and extrapolating the effective exponent to $L\to\infty$ yields
the estimate $\gamma \approx 1.7$.  The scaled genus distributions at long
times for $L=54$, $76$, and $90$ also show excellent data collapse
(Fig.~\ref{fig:genus-dist}).

\begin{figure}[h]
\includegraphics[width=0.2\textwidth]{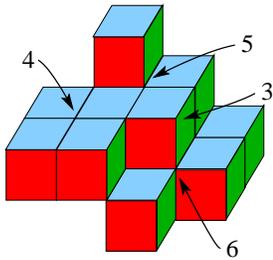}
\caption{\small A portion of an interface to illustrate vertices (labelled) that are
  shared among 3, 4, 5, or 6 edges.}
\label{vertices}
\end{figure}

Although we separately studied the energy and the genus, these two quantities
can be intimately connected by simple topological considerations.  We start
by simplifying Eq.~\eqref{Euler}.  For a closed surface that defines the
interface between spin domains, each face has 4 edges and each edge must be
shared between 2 adjacent faces.  Hence
\begin{equation}
\label{EF}
\mathcal{E}=2\mathcal{F}.
\end{equation}
Similarly, each edge has 2 vertices and each vertex is shared among 3, 4, 5,
or 6 adjacent edges (Fig.~\ref{vertices}).  This leads to the bounds
\begin{equation}
\label{VE}
\frac{\mathcal{E}}{3}\leq \mathcal{V}\leq \frac{2\mathcal{E}}{3}~.
\end{equation}  
Using these two relations in Eq.~\eqref{Euler} gives
\begin{equation}
-\frac{\mathcal{F}}{3}\leq\chi\leq \frac{\mathcal{F}}{3}~.
\end{equation}  
While the lower bound is useful, the upper bound can be replaced by the much
stronger condition $\chi\leq 2$, with the maximal value of 2 being achieved
for a sphere.  With this replacement, we obtain the following bounds on the
genus:
\begin{equation}
0\leq g\leq \frac{\mathcal{F}}{6}+1.
\end{equation}
However, $\mathcal{F}$ is directly related to the total energy of the system
because each face corresponds to a single pair of anti-aligned spins per spin
and thus has an energy cost of $+2$ (when the interaction strength is set to
1, see Eq.~\eqref{Ising}).  Thus $\mathcal{F}= L^3 E_L$.

Finally, we make the guess that the actual value of $g$ lies roughly midway
between the upper and lower bounds; namely, $g\propto \mathcal{F}$.  Using
our exponent definitions $E_L \sim L^{-\epsilon}$ and $\langle g\rangle\sim
L^\gamma$, our argument leads to the exponent inequality
\begin{equation}
\label{ineq}
\epsilon+\gamma\leq 3.
\end{equation}
Our numerical estimates for these two exponents given above, $\epsilon\approx
1$ and $\gamma\approx 1.7$ are consistent with Eq.~\eqref{ineq}.

We can carry this analysis a bit further by making use of some simple facts
in discrete differential geometry~\cite{GB}.  Let us define the number
of vertices with $m$ incident edges as $\mathcal{V}_m$.  It is useful to
introduce the notion of a ``defect'' that is associated with each vertex.
The defect for a vertex is defined as the difference between the sum of the
angles of all the faces at the vertex and $2\pi$.  With this definition, it
is easy to see that the defect of a vertices of types 3, 4, 5, and 6 are
$\frac{\pi}{2}$, 0, $-\frac{\pi}{2}$, and $-\pi$, respectively.  For any
domain, the total defect of all vertices on the surface equals $2\pi\,\chi$;
this is essentially the Gauss-Bonnet theorem for a discrete
interface~\cite{GB}.  Thus we have the general relation (see
Eq.~\eqref{Euler})
\begin{equation}
\frac{\pi}{2}\,\mathcal{V}_3  - \frac{\pi}{2}\, \mathcal{V}_5  - \pi\,\mathcal{V}_6
= 2\pi\,\chi= 4\pi(1-g).
\end{equation}
Therefore the genus of a surface and the number of vertices of various types
are related by
\begin{equation}
g= 1 +\frac{1}{8}(2\mathcal{V}_6+\mathcal{V}_5-\mathcal{V}_3)\,.
\end{equation}
From the examples of domains shown throughout this work, almost all vertices
are of type $m=4$ while vertices of type $m=5$ seem to be next most common.
Vertices of type $m=3$ are associated with blinkers and therefore should be
few in number.  Vertices of type $m=6$ arise at a 3-fold branch of the domain
and therefore seem to be the most rare.  Thus we expect that vertices of type
$m=5$ scale the same way as $\langle g\rangle$ which numerically appears to
grow as $L^{1.7}$.

\subsection{Survival Probability}
\label{survival}

The relaxation process is naturally characterized by $S(t)$, the probability
that the energy of the system is still decreasing at time $t$.  Since
energy-lowering spin flips occur rarely at long times, it is not immediately
evident whether the most recent energy-lowering spin flip is the last such
event or whether another energy-lowering flip event will occur sometime in
the distant future.  To determine if the energy has reached its final value
in an efficient way, we use the following algorithm that is a variant of our
acceleration algorithm.  As a preliminary, we separately track both
positive-energy and zero-energy flippable spins; the former are those for
which the energy decreases if such a spin actually flips.  We start the
simulation by running zero-temperature Glauber dynamics until no
positive-energy flippable spins remain.  At this time, defined as $T_0$, the
configuration $\mathcal{C}_0$ may have reached the final value of the energy.

\begin{figure}[ht]
\includegraphics[width=0.35\textwidth]{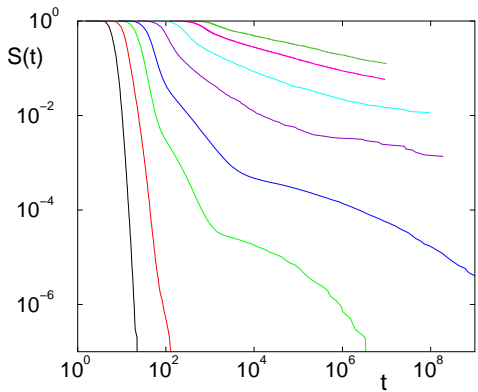}
\caption{\small Survival probability $S(t)$ versus time $t$ for
  $L=4,6,8,10,14,20, 30$ and $40$ (lower left to upper right) on a double
  logarithmic scale.  The averaging has been performed over $10^7$
  realizations for $L \leq 10$, $10240$ realizations for $L= 14, 20, 30$, and
  2048 realizations for $L=40$.}
\label{fig:S}
\end{figure}

\begin{figure}[ht]
\includegraphics[width=0.35\textwidth]{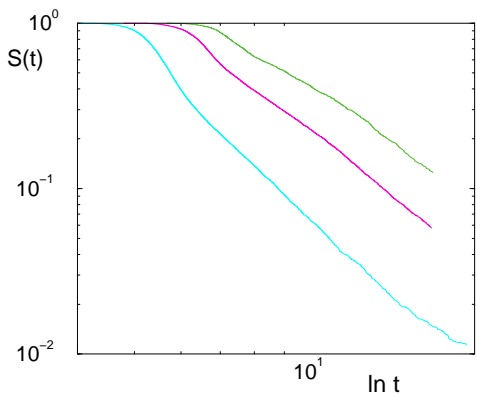}
\caption{\small $S(t)$ versus $\ln t$ for $L=20, 30,$ and 40 on a double logarithmic
  scale.}
\label{fig:S-large}
\end{figure}

We now proceed as follows:
\begin{itemize}
\itemsep -0.5ex
\item Apply an infinitesimal magnetic field.  If an energy-lowering spin-flip
  occurs, then the energy of $\mathcal{C}_0$ is not the final energy.  In
  this case, the system is returned to $\mathcal{C}_0$ and subsequently
  evolves by Glauber dynamics until again no positive-energy spins remain and
  a new candidate final configuration and final time is reached.

\item If an energy-lowering spin-flip does not occur, the system is returned
  to $\mathcal{C}_0$ and a field is applied in the opposite direction.
  Again, if an energy-lowering spin-flip occurs, the system is returned to
  $\mathcal{C}_0$ and subsequently evolves by Glauber dynamics until no
  positive-energy spins remain and a new candidate final state is reached.

\item If an energy-lowering spin-flip does not occur after the field has been
  applied in both directions, then $\mathcal{C}_0$ is at the final energy and
  the survival time equals $T_0$.

\end{itemize}

We record the time when the energy of a system stops changing, from which we
infer the time dependence of the survival probability $S(t)$.  This time
dependence is both surprisingly complex and extremely slow for small system
sizes (Fig.~\ref{fig:S}).  For example, for $L=10$, $40$ realizations had not
yet relaxed to their ultimate energy by $t=10^9$, a time that is seven orders
of magnitude beyond the coarsening time scale of $10^2$.  The data cutoff for
$L\geq 10$ was imposed because of CPU time limitations.  By $L=20$, the
dependence of $S(t)$ on $t$ becomes smooth and reasonably systematic and a
plot of $S(t)$ versus $\ln t$ on double logarithmic scale
(Fig.~\ref{fig:S-large}) suggests that the long-time data can be reasonably
fit to an inverse logarithmic dependence $S(t)\sim (\ln t)^{-\sigma}$, with
$\sigma\approx 3$.

\section{Discussion}
\label{summary}

We investigated the evolution of the kinetic Ising model that is endowed with
single spin-flip dynamics on a finite cubic lattice with periodic boundary
conditions.  The system starts in the antiferromagnetic state and is quenched
to zero temperature.  The details of the initial conditions are secondary as
long as the magnetization vanishes.  (If the initial magnetization is
non-zero, the evolution is much simpler and the Ising ferromagnet falls into
a ground state.)~ We asked the simple question: what happens?  A natural
expectation might be that the ground state should be reached.  A more
comprehensive version of this presumption is encapsulated by the {\em central
  dogma} of coarsening, which asserts:
\begin{enumerate}
\item Ising ferromagnets have just two metastable states, which coincide with
  the ground states.
\item If an Ising ferromagnet is endowed with zero-temperature single
  spin-flip dynamics, or more generally with a non order-parameter conserving
  dynamics, then one of the two ground states is necessarily reached.
\item The time to reach a ground state scales with the linear dimension of
  the system as $L^2$.
\end{enumerate}

This central dogma is indeed correct in one dimension.  However, for
two-dimensional Ising ferromagnets, there are numerous metastable states that
consist of single-phase stripes whose total number grows as
$\mathcal{M}_2\sim g^L$ \cite{SKR01}, where $g=\tfrac{1}{2}(\sqrt{5}+1)$ is
the golden ratio.  Nevertheless, the failure of the central dogma in two
dimensions is rather benign, as one of the ground states is reached
\cite{SKR01,SKR02,BKR09} with probability close to 2/3.  Moreover, for most
realizations, the final state (either a ground state or a stripe state) is
approached in a time that scales as $L^2$.

In three dimensions, however, the central dogma completely fails; viz., all
its three basic tenets are wrong.  First, the number of metastable states
$\mathcal{M}_3$ scales exponentially with the system size:
$\ln\mathcal{M}_3\sim L^3$ \cite{AP}.  Further, the ground states are never
reached (for sufficiently large systems) and the relaxation time is
anomalously long.  We provided heuristic and numerical evidence that the
relaxation time scales as $e^{L^2}$.  Thus for a macroscopic system with
$L\sim 10^8$ the relaxation time considerably exceeds any time scale in the
Universe.

Since the approach to the long-time state is extraordinarily slow, even for a
system as small as $10\times 10\times 10$, there are still realizations
(albeit a small fraction) for which the energy has not yet reached its final
value by $t=10^9$, whereas the standard coarsening time is of the order of
$10^2$.  We constructed a physical picture, based on the coalescence of
blinker-like configurations, that instead predicts a relaxational time scale
that grows exponentially in the linear dimension of the system.  In
particular, the survival probability $S(t)$, defined as the probability that
the energy has not yet relaxed to its final value by time $t$ seems to decay
as a power law in $1/\ln t$.  While the mechanism of blinker coalescence
appears plausible, we do not have a theoretical explanation for the
functional form of $S(t)$.  The primary feature of the relaxation that we
wish to emphasize is that the standard picture of coarsening, characterized
by a time that scales as $L^2$, is inappropriate for the three-dimensional
Ising model with zero-temperature Glauber dynamics.

Another striking feature of the three-dimensional Ising ferromagnet is that a
set of connected metastable microstates with equal energy is reached rather
than a `frozen' metastable state.  Each such set of states contains a small
number number of flippable blinker spins that can flip {\it ad infinitum}
without any energy cost.  Thus the system can wander within one of these
iso-energy sets forever.  Even though the number of blinker spins is small,
the spatial volume over which the blinker spins can roam comprises of the
order of 10\% of the volume of system in the limit of large $L$.

The topology of the long-time state is much richer than that of the
corresponding two-dimensional system.  In two dimensions, the only possible
states at infinite time are the ground state or an even number of alternating
single-phase stripes (for periodic boundary conditions).  In contrast, the
long-time states in three dimensions are highly interpenetrating and contain
many holes (Fig.~\ref{fig:g-20}).  Correspondingly, the average genus of the
domains scales with linear dimension as $L^{1.7}$.  Aside from this global
characterization of the domains, it is not clear what are the most useful
measures of the domain geometry.

\begin{figure}[ht]
\includegraphics[width=0.15\textwidth]{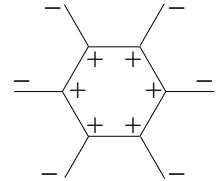}
\caption{\small Example of a frozen cluster of up spins on the hexagonal lattice.  }
\label{hexagon}
\end{figure}

While we focused on the cubic lattice, similar behaviors should arise for the
kinetic Ising model on other {\em even}-coordinated lattices in three
dimensions.  We deliberately avoided {\em odd}-coordinated lattices or other
complex networks where the coordination can be odd, as the zero-temperature
Ising-Glauber system quickly freezes, in disagreement with the central dogma
predictions.  However, this freezing has a local and trivial nature.  For
example, for an odd-coordinated lattice, single-phase droplets can arise, in
which spins within a droplet each have more internal than external
neighbors~\cite{NS,pontus,dean,CP06,H09}, as illustrated for the example of a
hexagonal droplet on the hexagonal lattice (Fig.~\ref{hexagon}).  These
droplets can thus remain forever in the phase opposite to that of the
background.  Another example of this local freezing is the Ising model with
zero-temperature Kawasaki (spin exchange) dynamics~\cite{book}, where again
local defects quickly arise that stop the overall relaxation process.

\begin{figure}[ht]
\includegraphics[width=0.15\textwidth]{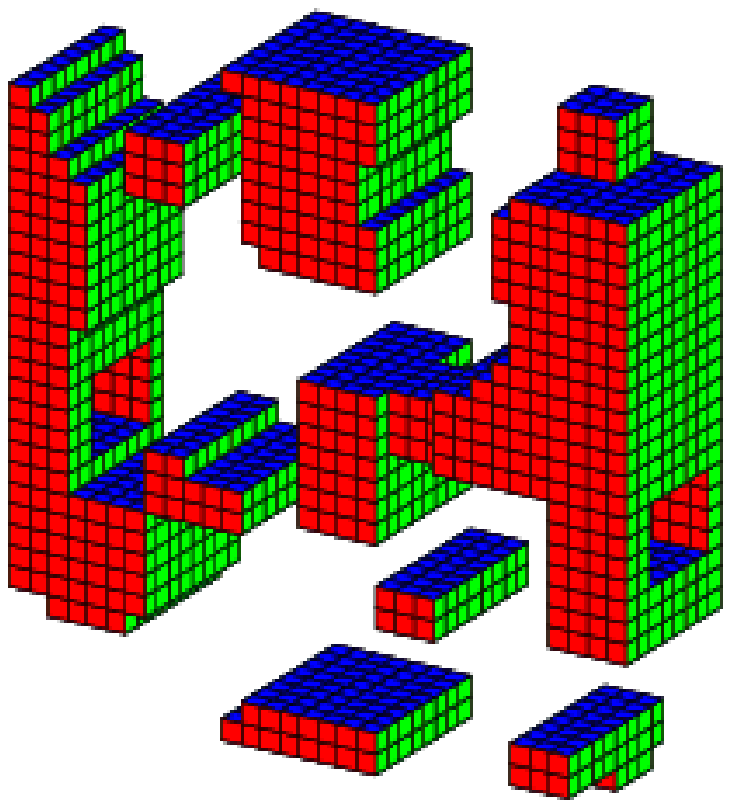}\includegraphics[width=0.15\textwidth]{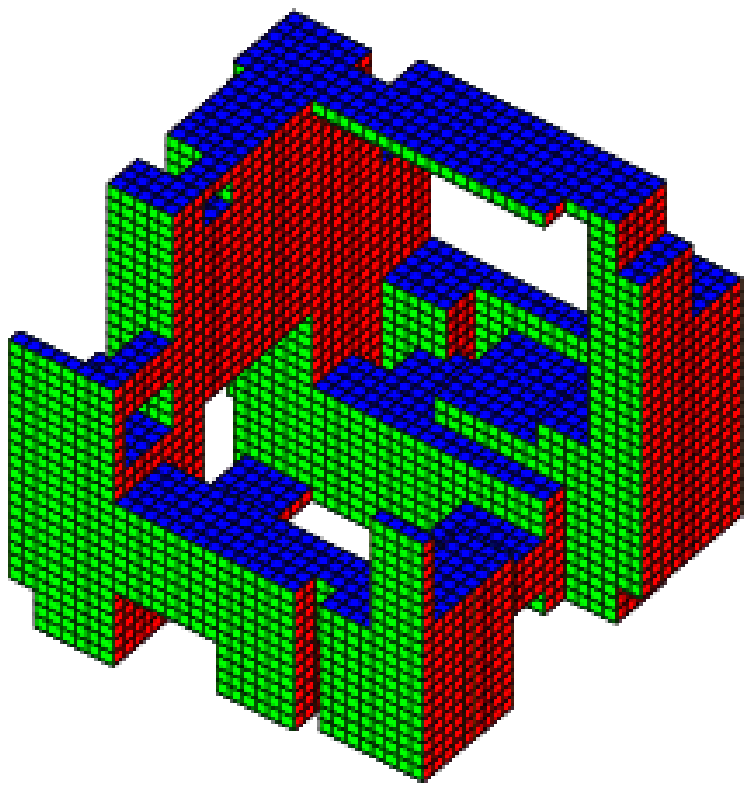}\includegraphics[width=0.15\textwidth]{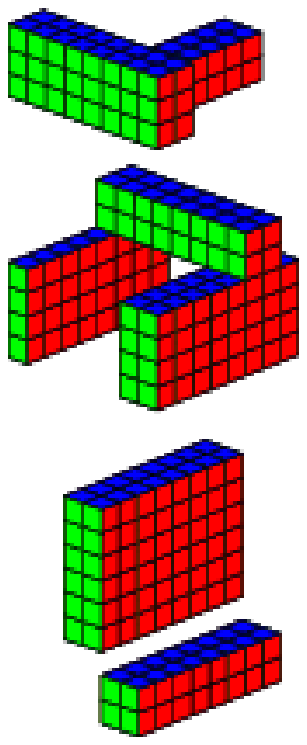}
\caption{\small Example long-time states for a $32\times 32\times 8$ slab, a
  $32^3$ cube, and a $8\times 8\times 32$ bar.}
\label{non-cubic}
\end{figure}

Intriguing and mostly unexplored behaviors arise for non-cubic systems
(Fig.~\ref{non-cubic}), for example, a $L\times L\times aL$ system.  When the
aspect ratio $a$ is small, the system becomes a thin square slab.  We are
generically interested in the thermodynamic limit $L\to \infty$ with $a$
fixed, so a thin square slab does not reduce to a two-dimensional system.
When the slab is thin, the long-time state resembles Swiss cheese, with
directed holes perpendicular to the slab.  Because of the periodic boundary
condition in all directions, there is no possibility of forming the stripe
states that arise in the two-dimensional system~\cite{SKR01,SKR02,SS,ONSS06}.
For $a\gg 1$, corresponding to a long bar, the long-time state consists of a
series of alternating domains of the two phases.  As the bar becomes wider,
percolation in the long direction eventually occurs, and the geometry begins
to resemble the plumber's nightmare (middle panel of Fig.~\ref{non-cubic}).

It is striking that the three-dimensional system is so much more complicated
than the corresponding two-dimensional case.  This fact suggests that many
more surprises await discovery for the kinetic Ising model in higher
dimensions.  Moreover, most of our findings are empirical in nature and they
beg for the development of new theoretical perspective and geometrical
descriptions of the domains.  Another challenging extension of the present
work is to describe the fate of spin systems with a non-scalar order
parameter, such as the XY model, that are quenched to zero temperature

\smallskip

We gratefully acknowledge financial support from NSF grant DMR0906504 (JO and
SR) and NSF grant CCF-0829541 (PLK).

\appendix

\section{Accuracy of the Acceleration Algorithm}
\label{sec:accel}

To test the accuracy of our acceleration algorithm, the system is evolved
until a cutoff time $\tau$ with zero-temperature Glauber dynamics.  Then at
time $\tau$, an infinitesimal magnetic field is applied, that alternates as
the system descends successive energy plateaux, until no flippable spins
remain.  Table~\ref{tab:compare} gives the final energy that is reached
(where no flippable spins remain) when the acceleration algorithm is applied
for different cutoff times $\tau$.  These data are shown for the cases
$L=10$, 20, and 100, with $10^7$, 10240, and 128 realizations, respectively.
The extremely weak dependence of the final energy as a function of $\tau$
indicates the level of accuracy of the acceleration algorithm.

\begin{table}[H]
\begin{center}
\begin{tabular}{|c|r|c|c|}
\hline
$L$ & $\tau$ & $\left<E_L\right>$ & $R$ \\ \hline
10 & 500 &  .4245900960 &  \\
 & $10^9$ &  .4245901020 &  34 \\
 & $10^{10}$  & .4245901020 &  60 \\ \hline
20 & 2000 & .283285352 &  \\
 & $10^5$ & .283287939 &  1.4 \\
 & $10^6$ & .283287939 & 5.2 \\
 & $10^7$ & .283287939 & 39 \\
 & $10^8$ & .283288232 &  378 \\ \hline
100 & $5\times 10^4$  & .083666469 & \\
 & $10^5$  & .083663406 &  1.2 \\
 & $5\times 10^5$  & .083662656 & 4.7 \\
 & $10^6$  & .083662594  & 7.2 \\
 & $10^7$ & .083662531  & 67 \\ \hline
\end{tabular}
\caption{Average final energies $\langle E_L\rangle$ for different cutoff
  times $\tau$ and system sizes $L$.}
\label{tab:compare}
\end{center}
\end{table}

The last column gives the ratio $R$ of the CPU time needed to simulate the
system until no flippable spins remain when the acceleration algorithm is
imposed at the cutoff time $\tau$ compared to imposing this algorithm at
$\tau= 5L^2$.  For example, it took 67 times longer to simulate the $L=100$
system by running zero-temperature Glauber dynamics to $t=10^7$ and
subsequently imposing the acceleration algorithm compared to running
zero-temperature Glauber dynamics to $t=5\times 10^4$ and then imposing the
acceleration algorithm.  The relative difference in the energies by the two
protocols is approximately $5\times 10^{-5}$, thus providing justification
for our use of the acceleration algorithm at $t=5L^2$.

\section{Small Systems, Blinker States, Number of Clusters}
\label{small}

The evolution of the smallest possible lattice $L=2$ helps to illustrate the
complexities of larger systems.  When $L=2$, there are 8 spins and $2^8=256$
possible states that can be enumerated to determine all details of the
evolution.  There are also only two possible final states: the ferromagnetic
ground state (F) and a static metastable state (M) that consists of a square
of four spins of one sign and an adjacent square of four spins of the
opposite sign.  There are nine distinct paths in state space that start at
the antiferromagnetic state and end at these two final states
(Table~\ref{tab:L=2}).  The average survival time until the system reaches
one of the final states is $\frac{221}{120}= 1.841666\ldots$, while the
probability of ultimately reaching the F final state is $\frac{11}{14}$.

\vskip -5ex
\begin{table}[H]
\begin{center}
\begin{tabular}{|c|c|r|c|c|}
\hline
path & \# flips & time~~ & prob. & final state\\
\hline
1 & 4 & $\frac{43}{56}$ & $\frac{2}{21}$ & M \\
2 & 4 & $\frac{143}{168}$ & $\frac{1}{14}$ & M\\
3 & 6 & $\frac{341}{280}$ & $\frac{1}{21}$ & M\\
4 & 4 & $\frac{85}{56}$ & $\frac{3}{14}$ & F\\
5 & 6 & $\frac{1583}{840}$ & $\frac{1}{35}$ & F\\
6 & 6 & $\frac{1793}{840}$ & $\frac{4}{35}$ & F\\
7 & 6 & $\frac{127}{56}$ & $\frac{4}{21}$ & F\\
8 & 6 & $\frac{395}{168}$ & $\frac{1}{7}$ & F\\
9 & 8 & $\frac{761}{280}$ & $\frac{2}{21}$ & F\\
\hline
\end{tabular}
\caption{The nine state-space paths to the final state for $L=2$ starting from 
  the antiferromagnetic initial condition.  Listed for each path are the number 
  of spin flips until the final state is reached, the time to reach the final 
  state on the path, and the probability of the path.  Also listed is the 
  final state for each path, either metastable (M) or ferromagnetic (F). }
\label{tab:L=2}
\end{center}
\end{table}

A complete enumeration is already not feasible for linear dimension $L=4$,
where the number of states is $2^{64}\approx 1.84\times 10^{19}$; thus
simulations are necessary when $L\geq 4$.  For $L=4$, the average survival
time is now 6.16, while the longest survival time observed in any realization
is 22.2.  For $L=6$ and 8, the respective average survival times are 11.9 and
27.9, while the longest survival times are 128 and $3.43\times 10^6$.  For
$L=10$, realizations that live beyond $t=10^{10}$ are possible, although
rare, and it is not possible to quote an average survival time.  The
existence of such long-lived realizations for $L\geq 10$ contributes to the
difficulty in the understanding of the behavior of the survival probability.

\begin{table}[H]
\begin{center}
\begin{tabular}{|c|c|c|c|}
\hline
$L$ & $P_g$ & $P_f$ & $P_b$ \\
\hline
2 & $\frac{11}{14}$ & $\frac{3}{14}$ & 0 \\
4 & 0.6814(1) & 0.3186(1) & 0 \\
6 & 0.3523(2) & 0.6353(2) & 0.01246(4) \\
8 & 0.1842(1) & 0.7373(1) & 0.07853(9) \\
10 & 0.1045(1) & 0.7170(1) & 0.1785(1) \\
20 & 0.01377(4) & 0.3059(1) & 0.6803(1) \\
32 & 0.00322(8) & 0.1091(4) & 0.8877(4) \\
54 & 0.00066(6) & 0.0406(4) & 0.9587(4) \\
76 & 0.00040(6) & 0.0250(5) & 0.9746(5) \\
90 & 0.00039(6) & 0.0199(4) & 0.9797(4) \\
\hline
\end{tabular}
\caption{Probabilities of reaching the ground state $P_g$, the frozen state $P_f$, and 
  the blinker state $P_b$  versus $L$.  The error in 
  the last digit is shown in parentheses.}
\label{tab:gfb}
\end{center}
\end{table}

It is also worth noting that infinitely long-lived blinker states first
appear for the case of $L=5$.  Table~\ref{tab:gfb} gives the data for the
probabilities of reaching the ground state, a frozen static state, or a
blinker state as a function of $L$.  The former two probabilities decrease
rapidly with $L$ and appear to approach zero for large $L$, while the
probability that a blinker state is reached approaches 1 as $L$ increases
(see also Fig.~\ref{blinker-prob}).  As mentioned at the beginning of
Sect.~\ref{long}, the long-time state almost always consists of exactly two
clusters.  Table~\ref{tab:nc} gives the probabilities that the long-time
state consists of one, two, three or more than three clusters as a function
of $L$.

\begin{table}[H]
\begin{center}
\begin{tabular}{|c|c|c|c|c|}
\hline
L & P(1) & P(2) & P(3) & P($>$3) \\
\hline
2 & $\frac{11}{14}$ & $\frac{3}{14}$ & 0 & 0 \\
4 & .6814(1) & .3186(1) & 0 & $<$.0000001 \\
6 & .3523(2) & .6475(2) & .000245(5) & $<$.0000001 \\
8 & .1842(1) & .8128(1) & .00303(2) & .0000004(2) \\
10 & .1045(1) & .8866(1) & .00893(3) & .000015(1) \\
20 & .01377(4) & .96052(6) & .02475(5) & .00096(1) \\
32 & .00322(8) & .9720(2) & .0230(2) & .00180(6) \\
54 & .00066(6) & .9802(3) & .0171(3) & .0020(1) \\
76 & .00040(6) & .9824(4) & .0150(4) & .0022(1) \\
90 & .00039(6) & .9839(4) & .0133(4) & .0024(2) \\
\hline
\end{tabular}
\caption{Probabilities of reaching a state that contains 1, 2, 3, or $>3$ 
  clusters at long times, with the error on 
  the last digit in parentheses.}
\label{tab:nc}
\end{center}
\end{table}


\begin{thebibliography}{99}
\bibitem{2d} In contrast, final states are topologically trivial in two
  dimensions; the only possibilities are the ground state or a static
  metastable state that consists of an even number of straight stripes.
\end{thebibliography}
\end{document}